\documentclass[12pt]{article}
\usepackage{graphicx}
\usepackage{enumerate}
\usepackage{natbib}
\usepackage{url} 
\usepackage{bbm}
\RequirePackage{amsthm,multirow,amsmath,fancybox,amssymb}
\usepackage[doublespacing]{setspace}
\usepackage{float}
\usepackage{subcaption}

\usepackage{xr}
\usepackage{cleveref}

\makeatletter
\newcommand*{\addFileDependency}[1]{
  \typeout{(#1)}
  \@addtofilelist{#1}
  \IfFileExists{#1}{}{\typeout{No file #1.}}
}
\makeatother
\newcommand*{\myexternaldocument}[1]{%
    \externaldocument{#1}%
    \addFileDependency{#1.tex}%
    \addFileDependency{#1.aux}%
}

\myexternaldocument{supp}

\usepackage{algorithm}
\usepackage{algorithmic}

\newcommand{\blind}{1}

\begin{document}

\def\spacingset#1{\renewcommand{\baselinestretch}%
{#1}\small\normalsize} \spacingset{1}


\if1\blind
{
  \title{\bf An Efficient Two-Dimensional Functional Mixed-Effect Model Framework for Repeatedly Measured Functional Data}
  \author{CHENG CAO \\
    Department of Data Science, City University of Hong Kong \\
    and \\
    JIGUO CAO \\
    Department of Statistics and Actuarial Science, Simon Fraser University \\
    and \\
    HAO PAN,  YUNTING ZHANG, FAN JIANG \\
    Department of Developmental and Behavioral Pediatrics, \\ Shanghai Children’s Medical Center, \\School of Medicine, Shanghai Jiao Tong University \\ 
    and \\
    XINYUE LI  \thanks{Corresponding author: Xinyue Li, Email: xinyueli@cityu.edu.hk}\hspace{.2cm}\\
    Department of Data Science, City University of Hong Kong}
  \maketitle
} \fi

\if0\blind
{
  \bigskip
  \bigskip
  \bigskip
    \title{\bf An Efficient Two-Dimensional Functional Mixed-Effect Model Framework for Repeatedly Measured Functional Data}
  \maketitle
  \medskip
} \fi

\bigskip
\begin{abstract}

With the rapid development of wearable device technologies, accelerometers can record minute-by-minute physical activity for consecutive days, which provides important insight into a dynamic association between the intensity of physical activity and mental health outcomes for large-scale population studies. Using Shanghai school adolescent cohort we estimate the effect of health assessment results on physical activity profiles recorded by accelerometers throughout a week, which is recognized as repeatedly measured functional data. To achieve this goal, we propose an innovative two-dimensional functional mixed-effect model (2dFMM) for the specialized data, which smoothly varies over longitudinal day observations with covariate-dependent mean and covariance functions. The modeling framework characterizes the longitudinal and functional structures while incorporating two-dimensional fixed effects for covariates of interest. We also develop a fast three-stage estimation procedure to provide accurate fixed-effect inference for model interpretability and improve computational efficiency when encountering large datasets. We find strong evidence of intraday and interday varying significant associations between physical activity and mental health assessments among our cohort population, which shed light on possible intervention strategies targeting daily physical activity patterns to improve school adolescent mental health. Our method is also used in environmental data to illustrate the wide applicability. Supplementary materials for this article are available online.

\end{abstract}

\noindent%
\vfill

\newpage
\spacingset{1.9} 
\section{Introduction}
\label{sec:intro}

A growing amount of research suggests an essential relationship between adolescents’ physical activity and mental health. For instance, prolonged sedentary behaviors have been found to elevate the risk of depression (\citealp{Poitras2016, Chaput2020, Tremblay2017}). Recent studies have further discovered that the context of when and where physical activity occurs are also influential factors (\citealp{Carson2016, Schmidt2017, da2022b}). Wearable devices can continuously record an individual's physical activity profiles over consecutive days, which offers unprecedented opportunities to analyze the varying associations with health outcomes for large-scale population studies (\citealp{morris2006a, morris2006b, xiao2015, park2018}). However, the dynamicity of the association is little studied as it relies on both functional variations of repeatedly measured activity profiles characterizing temporal differences across different times of the day and longitudinal variations capturing different days of the week. The development of new functional methodological tools for repeatedly measured functional data is required to address this scientific problem.

\begin{figure}[!t]
\centering
    \includegraphics[width=0.6\textwidth, height=0.375\textwidth]{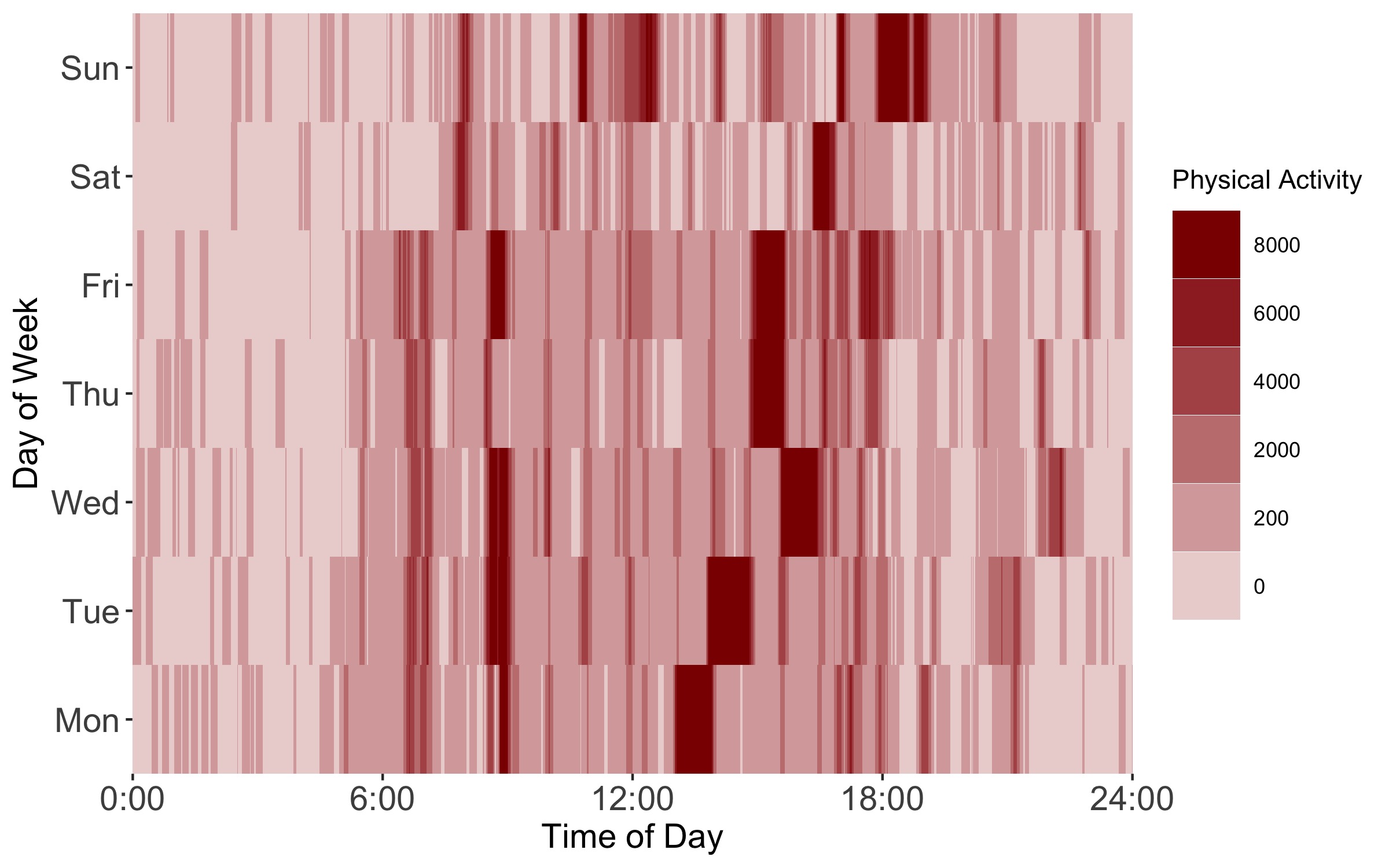}
\caption{Heatmap of a Shanghai school adolescent's activity profile showing minute-by-minute physical activity counts for a week. Data were obtained from wrist Actigraph.}
\label{fig:heatmap}
\end{figure}

Our motivating dataset comes from a Shanghai school adolescent study, which aims to examine whether a student's physical activity pattern is associated with mental health outcomes after adjusting for covariates such as demographic information. In this study, all adolescent participants wore ActiGraph for consecutive seven days to obtain activity count signals. A total of 2,313 students aged between 11 and 18 years from several schools in Shanghai participated in the study. The recorded accelerometer signals were summarized into minute-by-minute activity counts, resulting in a total of 1,440 functional grids over a day, and one example of the physical activity profile for one participant is shown in Figure \ref{fig:heatmap}. For each subject, we collected his/her demographic information and mental health assessment results. More details will be provided in Section \ref{sec:app}.

The physical activity recorded by advanced accelerometers can be seen as repeatedly measured functional observations with covariate-dependent mean and covariance functions. Understanding the intraday and interday relationships between physical activity and key factors can provide insights into intervention strategies for school adolescent health. However, the specialized data structure makes it difficult to characterize using one-dimensional models alone, which implies the importance of two-dimensional modeling to adequately capture the data natures. Furthermore, many methods struggle with scalability even when encountering smaller sample sizes and functional sampling grids than our motivating data (\citealp{cui}). To improve the computational efficiency of model estimation and inference is also challenging for practical application. 

Currently, repeatedly measured physical activity data monitored by wearable devices are commonly studied as one special case of longitudinal functional data analysis using functional mixed effect model (FMEM) (\citealp{morris2006a, zhu2019, cui, li2022}). The method aims at fixed-effects estimation and inference under the mixed-effect framework by applying splines (\citealp{guo, yuan, zhu2019}), functional principal component analysis (FPCA) (\citealp{goldsmith2015}), or Bayesian methods (\citealp{morris2006a,morris2006b, goldsmith2016}). While longitudinal functional data analysis mainly focuses on sparse and irregular longitudinal-based design (\citealp{greven, li2022}), repeatedly measured functional data are often densely sampled and measured over regular longitudinal visits. Hence, the analysis of repeatedly measured functional data is of independent interest (\citealp{chen2012, chen2017}), where the methodology can be refined in three aspects when taking into account the practical data structure. 

First, the variations of effect along both longitudinal and functional directions underscore the need for evaluation from two-dimensional rather than one-dimensional effects as in conventional longitudinal functional studies. While some research suggests bivariate functional models for symmetrical two-dimensional functional data, i.e. images, the modeling framework and estimation procedures rely essentially on their univariate cases and lead to extra computational burden (\citealp{zhang2011, ivanescu, morris2011}). Besides, they also inadequately account for complex four-dimensional correlation structures presented in longitudinal and functional contexts. Second, enhancing the correlation structure is crucial for improving model flexibility and universality. The random effect component of traditional one-dimensional FMEM incorporates longitudinal visits through a linear framework with additive assumptions (\citealp{chen2012, park2015, li2022}). In comparison, a nonparametric four-dimensional covariance function that considers continuity along two directions and enables a flexible model structure with minimal assumptions is a potential alternative. Third, there is a growing need to boost the computational efficiency of existing methods to apply to population studies. They are often computationally expensive due to challenges in modeling the complex four-dimensional correlation structure and conducting fixed-effect inference. Although some techniques have been proposed to enhance computational efficiency (\citealp{cui, li2022}), implementing FMEM in large-scale population studies remains challenging.

In this study, we aim to propose a novel two-dimensional functional mixed-effect model (2dFMM) framework for repeatedly measured functional data. Our model considers two-dimensional fixed effects for the specialized dense and regular longitudinal designs and leverages a nonparametric four-dimensional covariance structure. This flexible structure can effectively avoid covariance structure assumption violations that may often occur in longitudinal settings. We also develop a fast three-stage estimation procedure using pointwise and smoothing techniques for accurate bivariate coefficient function estimation, and it can effectively achieve good model interpretability and at the same time ensure fast computation. Moreover, the fixed-effect inference is built under the weak separability assumption of spatiotemporal covariance function to decompose marginal effects (\citealp{huang2017, lynch, liang2022}) and enable efficient estimation of the four-dimensional covariance function, leveraging FPCA and basis splines to relieve the computational burden. 

The rest of this paper is organized as follows. In Section \ref{sec:method}, we propose 2dFMM with estimation and inference procedure. Asymptotic results of the proposed estimator are also provided. Extensive simulation studies are conducted in Section \ref{sec:simu} to evaluate the performance of 2dFMM and compare it with existing approaches. Section \ref{sec:app} further applies 2dFMM to the Shanghai school adolescent study and Australia electricity demand data. Conclusion and discussion are in Section \ref{sec:conc}.

\section{Methods}
\label{sec:method}

\subsection{Two-Dimensional Functional Mixed-Effect Model}
\label{sec:2dFMMmodel}

The functional response is denoted by $Y_{i}(s,t)$, $i$th subject's profile functional time $t \in \mathcal{T}$, repeatedly measured at longitudinal time $s \in \mathcal{S}$, where $i=1,\dots,N$. We assume that the covariate-dependent mean function $\mu(s,t,\mathbf{X}_{i})$ is the linear combination of $P$-dimensional time-invariant or time-varying covariates of interest $\mathbf{X}_{i} = (1, x_{i,1}(s,t), \dots, x_{i,P}(s,t))^{T}$ at time $s$ and $t$, which is also known as a standard linear concurrent model. The proposed two-dimensional functional mixed-effect model (2dFMM) is 
\begin{equation}
\begin{aligned}
    Y_{i}(s,t) &= \mu(s,t,\mathbf{X}_{i}) + \eta_{i}(s,t) + \epsilon_{i}(s,t) \\
    &= \beta_{0}(s,t) + \sum_{p=1}^{P}x_{i,p}(s,t)\beta_{p}(s,t) + \sum_{j=1}^{\infty}\xi_{i,j}(t)\psi_{j}(s) + \epsilon_{i}(s,t),
\end{aligned}
\label{model:2dFMM}
\end{equation}
where $\{\beta_{0}(s,t), \beta_{1}(s,t), \dots, \beta_{P}(s,t)\}$ are corresponding coefficient functions, $\eta_{i}(s,t)$ is a bivariate random process with mean zero and four-dimensional covariance function $C(s,t;u,v)$, while random measurement error process $\epsilon_{i}(s,t)$ is mean zero with some covariance function $\mathcal{E}(s,t;u,v)$ and independent of $\eta_{i}(s,t)$. 

For dimension reduction,  $\eta_{i}(s,t)$ is typically decomposed via well-developed two-dimensional FPCA and represented in Karhunen-Loève expansion with orthonormal basis of $L^{2} (\mathcal{S}\times\mathcal{T})$. It ensures that a large proportion of the four-dimensional covariance function $C(s,t;u,v)$ can be explained by the top several terms, but is computationally demanding and not suitable for asymmetric bivariate arguments which are usually assumed in repeatedly measured functional data. To address these issues, we exploit the marginal covariance functions technique to separate the bivariate process for efficiency (\citealp{park2015, chen2017, li2022}). In model \eqref{model:2dFMM},  we define $W_{i}(s,t) = \eta_{i}(s,t) + \epsilon_{i,1}(s,t)$, where $\epsilon_{i}(s,t) = \epsilon_{i,1}(s,t) + \epsilon_{i,2}(s,t)$ for model identifiability;  $C_{\mathcal{W}}(s,u) = C_{\mathcal{S}}(s,u) + \Gamma(s,u)$, where $C_{\mathcal{S}}(s,u) = \int_{\mathcal{T}}C(s,t;u,t)dt$ is the marginal covariance with respect to $\mathcal{S}$ and . Let $\psi_{j}(s)$ be eigenfunction of marginal covariance function, such that $C_{\mathcal{W}}(s,u)=\sum_{j=1}^{\infty}\tau_{j}\psi_{j}(s)\psi_{j}(u)$; $\{\xi_{i,j}(t): j \geq 1\}$ are random coefficient functions obtained by the projection of $\eta_{i}(\cdot,t)$ subject onto the direction $\psi_{j}(s)$, i.e. $\langle W(\cdot,t), \psi_{j} \rangle_{\mathcal{S}} - \langle \epsilon_{1,i}(\cdot,t), \psi_{j} \rangle_{\mathcal{S}}$. It is easy to see that $\mathbb{E}[\xi_{i,j}(t)] = 0$ for $t \in \mathcal{T}$ and $\mathbb{E}[\langle \xi_{i,j}, \xi_{i,h}\rangle_{\mathcal{T}}] = \tau_{j}\mathbbm{1}_{\{j=h\}}$. We further define the covariance function $\Theta_{j}(t,v) = \mathbb{E}[\xi_{i,j}(t)\xi_{i,j}(v)]$, which depends on $j$th eigenfunction of marginal covariance $C_{\mathcal{S}}(s,u)$. Therefore, for any $s \neq u$ and $t \neq v$, we can obtain an expression of the four-dimensional covariance function as follows, 
\begin{equation}
\begin{aligned}
    C(s,t;u,v) &= \mathbb{E}[(Y_{i}(s,t) - \mu(s,t, \mathbf{X}_{i}))(Y_{i}(u,v) - \mu(u,v, \mathbf{X}_{i}))] \\
    &= \sum_{j=1}^{\infty}\psi_{j}(s)\psi_{j}(u)\Theta_{j}(t,v).
\end{aligned}
\label{eq:4dim}
\end{equation}

Moreover, regarding two components of measurement error $\epsilon_{i}(s,t)$,
$\epsilon_{i,1}(s,t)$ with covariance function $\Gamma(s,u)\mathbbm{1}_{\{t=v\}}$ depicts variation of repeated visits and white noise $\epsilon_{i,2}(s,t)$ is mean zero with variance $\varphi^{2}$. Therefore, $\mathcal{E}(s,t;u,v) = \{\Gamma(s,u) + \varphi^{2}\mathbbm{1}_{\{s=u\}}\}\mathbbm{1}_{\{t=v\}}$, where $\mathbbm{1}_{\{\cdot\}}$ is an indicator function.

Thus, the functional response $Y_{i}(s,t)$ is \textit{i.i.d.} Gaussian random process with mean $\mathbb{E}[Y_{i}(s,t)] = \mu(s,t, \mathbf{X}_{i})$ and variance function of $Y_{i}(s,t)$ is $\sigma^{2}(s,t)$, while for any $s \neq u$ and $t \neq v$, covariance function is $\sigma(s,t;u,v) = C(s,t;u,v) + \mathcal{E}(s,t;u,v)$.




\textit{Remark 1}. The proposed bivariate model differs from existing FMEM in two key aspects. First, the expectation of the functional response in our model depends on bivariate coefficient functions, which inherently imply possible variations over the longitudinal domain. Second, our model captures complex correlation structure through the four-dimensional covariance function $C(s,t;u,v)$, removing the constraint of the linear framework in FMEM and allowing a more flexible representation. In addition, our bivariate model addresses the pre-specification difficulty of FMEM design: whether using a random slope effect depends on hard-to-verify \textit{a priori} assumptions about longitudinal correlation structure. Our data-driven approach circumvents this specification difficulty. We demonstrate this better accommodates subtle complexities like distinctions between in-school weekdays and off-school weekend patterns for our motivating examples. 

\textit{Remark 2}. The four-dimensional covariance function in Equation \ref{eq:4dim} is decomposed under a weaker assumption than strong separability, which is defined as $C(s,t;u,v) = C_{\mathcal{S}}(s,u)C_{\mathcal{T}}(t,v)$ (\citealp{lynch}). We approximate $C_{\mathcal{T}}(t,v)$ within each score function $\xi_{i,j}(t)$ and corresponding covariance function $\Theta_{j}(t,v)$. The representation relaxes the restriction to the scope of strong separability and is theoretically sound because it delivers near-optimality under the appropriate assumptions (\citealp{chen2017, aston2017}). 

\textit{Remark 3}. Compared with FMEM in longitudinal functional studies, our coefficient functions are expanded from curves to surfaces, indicating two-dimensional effects on functional responses. It is deemed necessary for repeatedly measured functional data varying in two domains. Despite this, we can still approximate the univariate effect by taking the average of the bivariate coefficient function along either domain. 

Suppose that the sampling points $\{s_{r}, r=1,\dots,R\}$ and $\{t_{l}, l=1,\dots,L\}$ are prefixed in this study with design densities $f_{\mathcal{S}}(s)$ and $f_{\mathcal{T}}(t)$ such that $\int_{s_{1}}^{s_{r}} f_{\mathcal{S}}(s)ds  = r/R$ for $r \geq 1$ and similarly $\int_{0}^{t_{l}} f_{\mathcal{T}}(t)dt  = l/L$ for $l \geq 1$. Both densities are continuous second-order differentiable with the support $\mathcal{S}$ or $\mathcal{T}$, which are uniformly bounded away from zero and infinity. Suppose $\beta_{p}(s,t) = \beta_{p}(s)\beta_{p}(t)$ and only take account into $\mathcal{T}$, the satisfication of identification conditions $\mathbb{E}[\beta_{p}(s)] = 1$ implies $\beta_{p}(t) = \int_{\mathcal{S}}\beta_{p}(s,t)f_{\mathcal{S}}(s)ds$. Hence, given an estimator of bivariate coefficient function $\hat{\beta}_{p}(s,t)$, univariate function estimator can be obtained $\hat{\beta}_{p}(t) = R^{-1}\sum_{r=1}^{R}\hat{\beta}_{p}(s_{r},t)$
with covariance estimator $\widehat{\text{Cov}}\{\hat{\beta}_{p}(t), \hat{\beta}_{p}(v)\} =R^{-2}\sum_{r_{1}}\sum_{r_{2}}\widehat{\text{Cov}}\{\hat{\beta}_{p}(s_{r_{1}},t),\hat{\beta}_{p}(s_{r_{2}},v)\}$.

\subsection{Estimation of Model Components}
\label{sec:est}

We propose a computationally efficient three-stage model estimation procedure. First, we estimate the fixed effects using bivariate pointwise and post-smoothing (pointwise-smoothing) estimation. We also represent the four-dimensional covariance function $C(s,t;u,v)$ as a combination of eigenfunctions and B-spline basis functions for faster estimation. This is achieved by leveraging the decomposition of the marginal covariance. We assume densely and regularly sampled grids for both domains. Estimation with randomly sampled designs is discussed in Section \ref{sec:conc}. Our approach also lends itself well to parallelization, further accelerating the entire process.

Suppose design points $s_{r}$ and $t_{l}$ satisfy densities in \textit{Remark 3} , we let a pointwise response vector denoted by $\mathbf{Y}_{s_{r},t_{l}}$ and design matrix $\mathbf{X}_{s_{r},t_{l}} = (\mathbf{X}_{1, s_{r},t_{l}}^{T}, \dots, \mathbf{X}_{N,s_{r},t_{l}}^{T})^{T}$, where $\mathbf{X}_{i, s_{r},t_{l}} = (1, x_{i,1}(s_{r},t_{l}), \dots, x_{i,P}(s_{r},t_{l}))^{T}$. Our estimation procedure is as follows.

\textit{Step I: Bivariate Pointwise Estimation}. We reform a linear model 
$\mathbf{Y}_{s_{r},t_{l}} = \mathbf{X}_{s_{r},t_{l}}\boldsymbol{\beta}_{s_{r},t_{l}} + \mathbf{e}_{s_{r},t_{l}}$, where $\boldsymbol{\beta}_{s_{r},t_{l}} = (\beta_{0}(s_{r},t_{l}), \dots, \beta_{P}(s_{r}, t_{l}))^{T}$ and $\mathbf{e}_{s_{r}, t_{l}} \sim N(\boldsymbol{0}, \sigma^{2}_{s_{r}, t_{l}}\mathbf{I}_{N})$, under the mutual independence assumption, which is commonly adopted in other fixed-effect estimation approach (\citealp{fan2000, park2018, cui}). We use ordinary least squares estimator $\tilde{\boldsymbol{\beta}}_{s_{r},t_{l}} = (\mathbf{X}^{T}_{s_{r},t_{l}}\mathbf{X}_{s_{r},t_{l}})^{-1}\mathbf{X}^{T}_{s_{r},t_{l}}\mathbf{Y}_{s_{r},t_{l}}$ for initial estimation because it shares nice statistical properties and low computational cost. Considering the bivariate pointwise conditions, the covariance matrix estimates of coefficient functions can be obtained for any $s_{r_{1}}, s_{r_{2}}, t_{l_{1}}$, $t_{l_{2}}$,
\begin{equation}
 \text{Cov}\{\tilde{\boldsymbol{\beta}}_{s_{r_{1}},t_{l_{1}}}, \tilde{\boldsymbol{\beta}}_{s_{r_{2}},t_{l_{2}}}\} =\sigma_{s_{r_{1}},t_{l_{1}};s_{r_{2}},t_{l_{2}}}\mathbf{H}_{s_{r_{1}},t_{l_{1}}}\mathbf{H}^{T}_{s_{r_{2}},t_{l_{2}}},
 \label{eq:cov_pt_beta_2d}
\end{equation}
where $\mathbf{H}_{s_{r},t_{l}} = (\mathbf{X}^{T}_{s_{r},t_{l}}\mathbf{X}_{s_{r},t_{l}})^{-1}\mathbf{X}^{T}_{s_{r},t_{l}}$ and  the estimates for $\sigma_{s_{r_{1}},t_{l_{1}};s_{r_{2}},t_{l_{2}}}$ will be provided in step III later. Note that for pairwise observation $(s_{r},t_{l})$,
the standard error estimate of the linear model which is denoted by $\tilde{\sigma}^{2}_{s_{r},t_{l}}$ is the raw estimate to $\sigma^{2}_{s_{r},t_{l}}$.

\textit{Step II: Bivariate Smoothing}. As the raw estimate is obtained, bivariate smoothing is required to refine it by integrating neighboring temporal information. The post-smoothers for the bivariate process are rich, for example, bivariate P-splines (\citealp{eilers2003, marx2005}), thin plate regression splines (\citealp{wood2003}), tensor product smooths (\citealp{wood2006}), and sandwich (\citealp{xiao}). 

Here we illustrate the use of sandwich smoother due to its computational efficiency with nice asymptotic properties. Denote raw $p$th estimated bivariate coefficient function by matrix $\tilde{\boldsymbol{\beta}}_{p}= (\tilde{\beta}_{p}\big(s_{r}, t_{l})\big)_{R \times L}$, the refined estimator is simply expressed in a closed-form solution such that $\hat{\boldsymbol{\beta}}_{p} = \mathbf{S}_{2}\tilde{\boldsymbol{\beta}}_{p}\mathbf{S}_{1}$, where $\mathbf{S}_{1}$ and $\mathbf{S}_{2}$ are smoother matrices for $\mathcal{T}$ and $\mathcal{S}$ respectively, utilizing P-splines in different prespecified number of knots $K_{R}$ and $K_{L}$, respectively. 

With the help of the bivariate smoother, the variability of covariance estimator $\tilde{\sigma}^{2}_{s_{r},t_{l}}$ can be further diminished. Applying the sandwich smoother on standard error matrix $\widetilde{\mathbf{R}} = (\tilde{\sigma}^{2}_{s_{r},t_{l}})_{R \times L}$ gives the final covariance estimator $\widehat{\mathbf{R}} = (\hat{\sigma}^{2}_{s_{r},t_{l}})_{R \times L}$. However, there is no guarantee that the resulting estimators are all non-negative. The issue can be handled by trimming the negative values at zero (\citealp{yao2005, greven, cui}). 

In this study, we also provide tensor product smooths in the practical implementation of bivariate smoothing. To quantify the wiggliness of the raw estimator, sandwich smoother relies on differencing matrices to account for the distance between adjacent coefficients, while tensor product smooths use common second-order derivatives penalties. The performance of two smoother will be shown in detail in simulation studies.

\textit{Step III: Covariance Estimation}. To reduce the computing burden of the most time-consuming covariance estimation step, we employ flexible nonparametric methods incorporating FPCA on marginal covariance of $C_{\mathcal{S}}(\cdot, \cdot)$ and B-splines to approximate random functions $\xi_{i,j}(\cdot)$, tailored to imbalance number of grids for two domains. The efficient estimation procedure also consists of three stages. 


Firstly, we use the centered data, $\widetilde{Y}_{i}(s_{r},t_{l}) = Y_{i}(s_{r},t_{l}) - \hat{\mu}(s_{r},t_{l}, \mathbf{X}_{i})$, to estimate the marginal covariance function $C_{\mathcal{W}}(s,u)$, and obtain the estimates of eigenfunctions $\psi_{j}(s)$ and score functions $\xi_{i,j}(t)$. Specifically, given the refined estimator of coefficient functions from Step II, we pool $\{\widetilde{Y}_{i}(\cdot,t_{l}), i = 1,\dots N, l=1, \dots L\}$ and have sample covariance
\begin{equation*}
    \widetilde{C}_{\mathcal{W}}(s_{r_{1}},s_{r_{2}}) = (|\mathcal{T}|/NL)\sum_{i=1}^{N}\sum_{l=1}^{L}\widetilde{Y}_{i}(s_{r_{1}},t_{l})\widetilde{Y}_{i}(s_{r_{2}},t_{l}),
\end{equation*}
where $1 \leq r_{1} \leq r_{2} \leq R$. The storage memory and computation of $ \widetilde{C}_{\mathcal{W}}(s_{r_{1}},s_{r_{2}})$ is light because the number of longitudinal grids is relatively small. We obtain the resulting positive semi-definite covariance function estimates $\widehat{C}_{\mathcal{W}}(s_{r_{1}},s_{r_{2}})$ by kernel-based local linear smoothing in PACE algorithm, which excludes the diagonal entries due to the inflation by the effect of white noise (\citealp{yao2005}). Let $\{\hat{\psi}_{j}(s): j = 1,\dots, J\}$ be the estimated eigenfunctions and $\tilde{\xi}_{\mathcal{W},i,j}(t_{l}) = \int_{\mathcal{S}} \widetilde{Y}_{i}(s,t_{l})\hat{\psi}_{j}(s)ds$ be the estimated score function, which is used as the approximant to $\tilde{\xi}_{i,j}(t_{l})$. The number of components $J$ is determined by a fraction of variance explained (FVE), the threshold of which is set 0.99. Additionally, the variance of white noise $\varphi^{2}$ can be estimated as the average difference between $\widetilde{C}_{\mathcal{W}}(s_{r_{1}},s_{r_{2}})$ and $\widehat{C}_{\mathcal{W}}(s_{r_{1}},s_{r_{2}})$. 

Secondly, we estimate the marginal covariance functions $\Theta_{j}(t, v)$ by ``observed" functional data $\tilde{\xi}_{i,j}(t)$. Suppose each functional data has B-splines basis expansion $\tilde{\xi}_{i,j}(t_{l})=\mathbf{B}^{T}_{j}(t_{l})\mathbf{b}_{i,j}$, where $\mathbf{B}_{j}(t_{l}) = (B_{j1}(t_{l}), \dots, B_{jK}(t_{l}))^{T}$ and $\mathbf{b}_{i,j} = (b_{i,j1}, \dots, b_{i,jK})^{T}$,  $B_{jk}(t_{l})$ is $k$th B-splines basis function of $j$th principal component, $K$ is the number of basis functions. Let $K$ be the same for all $j$, the basis functions $\mathbf{B}(t_{l})$ do not rely on $j$. The covariance estimator of $\tilde{\xi}_{i,j}(t_{l})$, denoted by $\widehat{\Theta}_{j}(t_{l_{1}}, t_{l_{2}})$, can be obtained nonparametrically
\begin{equation*}
\begin{aligned}
    \widehat{\Theta}_{j}(t_{l_{1}}, t_{l_{2}}) = N^{-1}\sum_{i=1}^{N}\hat{\xi}_{i,j}(t_{l_{1}})\hat{\xi}_{i,j}(t_{l_{2}}) = N^{-1}\sum_{i=1}^{N}\mathbf{B}^{T}(t_{l_{1}})(\hat{\mathbf{b}}_{i,j}^{T}\otimes \hat{\mathbf{b}}_{i,j})\mathbf{B}(t_{l_{2}}).
\end{aligned}
\end{equation*}

The choice of the value of $K$ depends on a tradeoff between capturing variations adequately and ensuring computational efficiency. To ensure that the majority of variations are captured by large enough number of basis functions, we also consider the computational efficiency of the basis function expansion. Compared with FPCA, which has the computational complexity $O(NL^{2} + L^{3})$ for each $\tilde{\xi}_{i,j}(t)$  (\citealp{chung2021, cui2023}), the usage of B-splines requires $O(NLK^{2})$ computations. It implies that when $K < (L+L^{2}/N)^{1/2}$, our approach offers a lighter computational cost. Naturally, we can set $K = K_{L}$ and $K_{L} = \min\{L/2, 35\}$ is recommended by \cite{xiao}. Thus, in practical usage, we suggest $K_{L} = \min\{(L+L^{2}/N)^{1/2}, L/2, 35\}$.

Finally, given Equation \eqref{eq:4dim}, we obtain the estimator of four-dimensional covariance function $C(s_{r_{1}},t_{l_{1}};s_{r_{2}},t_{l_{2}})$ as follows. 
\begin{equation}
    \widehat{C}(s_{r_{1}},t_{l_{1}};s_{r_{2}},t_{l_{2}}) =N^{-1}\mathbf{B}^{T}(t_{l_{1}}) \Big\{\sum_{j=1}^{J}\hat{\psi}_{j}(s_{r_{1}})\hat{\psi}_{j}(s_{r_{2}})\sum_{i=1}^{N}\hat{\mathbf{b}}_{i,j}^{T}\otimes \hat{\mathbf{b}}_{i,j}\Big\}\mathbf{B}(t_{l_{2}}),
\label{eq:Chat}
\end{equation}
and therefore, $\hat{\sigma}_{s_{r_{1}},t_{l_{1}}; s_{r_{2}},t_{l_{2}}} = \widehat{C}(s_{r_{1}},t_{l_{1}}; s_{r_{2}},t_{l_{2}}) + \hat{\sigma}^{2}_{s_{r_{1}},t_{l_{1}}}\mathbbm{1}_{\{s_{r_{1}} = s_{r_{2}}, t_{l_{1}} = t_{l_{2}}\}}$. 

\subsection{Inference Procedure}

Here we show the construction of pointwise and simultaneous confidence bands. These inference procedures require the estimator of $\text{Cov}\{\tilde{\boldsymbol{\beta}}_{s_{r_{1}},t_{l_{1}}}, \tilde{\boldsymbol{\beta}}_{s_{r_{2}},t_{l_{2}}}\}$, which can be explicitly obtained by Equation \eqref{eq:cov_pt_beta_2d} and \eqref{eq:Chat}. However, this sample covariance function is wiggly because it relies on a pointwise estimator, although the B-splines smoothing technique controls the roughness to some degree. We still need a covariance estimator of refined coefficient $\text{Cov}\{\hat{\boldsymbol{\beta}}_{s_{r_{1}},t_{l_{1}}}, \hat{\boldsymbol{\beta}}_{s_{r_{2}},t_{l_{2}}}\}$ in the form of the sandwich smoother, which is coincidentally equivalent to the covariance smoothing approach (\citealp{xiao2016, xiao2018}). Let $\tilde{\beta}_{p} = \text{vec}(\tilde{\boldsymbol{\beta}}_{p})$ indicate a matrix is stacked by column and $\widehat{\text{Cov}}\{\tilde{\beta}_{p}, \tilde{\beta}_{p}\}$ be a $RL \times RL$ covariance matrix estimates of $p$th bivariate coefficient function formed by $p$th diagonal element of $\widehat{\text{Cov}}\{\tilde{\boldsymbol{\beta}}_{s_{r_{1}},t_{l_{1}}},\tilde{\boldsymbol{\beta}}_{s_{r_{2}},t_{l_{2}}}\}$ for all $s_{r_{1}}, s_{r_{2}}, t_{l_{1}}$, $t_{l_{2}}$. By tensor product properties $\hat{\beta}_{p} = (\mathbf{S}_{1} \otimes \mathbf{S}_{2})\tilde{\beta}_{p}$,  the ultimate four-dimensional covariance estimator of the $p$th bivariate coefficient function is 
\begin{equation}
    \widehat{\text{Var}}\{\hat{\beta}_{p}(s_{r},t_{l})\} = \boldsymbol{e}_{s_{r},t_{l}}^{T}(\mathbf{S}_{1} \otimes \mathbf{S}_{2})\widehat{\text{Cov}}\{\tilde{\beta}_{p}, \tilde{\beta}_{p}\}(\mathbf{S}_{1} \otimes \mathbf{S}_{2})^{T} \boldsymbol{e}_{s_{r},t_{l}}.
    \label{eq:cov_sm_beta_2d}
\end{equation} 
where $\boldsymbol{e}_{s_{r},t_{l}}$ denotes $RL$-dimensional unit vector with 1 at the $(l-1)L+r$th entry.

The analytic inference for two-dimensional fixed effects using confidence bands is straightforward. Depending on the pointwise variability for every pair of $(s_{r},t_{l})$ estimated by Equation \eqref{eq:cov_sm_beta_2d}, the $\pm 2$ standard error surfaces can be constructed by 
\begin{equation*}
    \hat{\beta}_{p}(s_{r},t_{l}) \pm 2\widehat{\text{Var}}\{\hat{\beta}_{p}(s_{r},t_{l})\}^{1/2}.
\end{equation*}
Note that our estimator is biased so that the standard error surface is also called a $95\%$ pointwise confidence bands (PCB) if neglecting the effect of bias term based on the nice approximation property of sandwich smoother (\citealp{fan2000, xiao, zhu2019}). While PCB is efficient to construct in analytical form, using it for statistical inference can be flawed because it ignores the inherent correlation of functional data and results in false positives (\citealp{crainiceanu2024}).

\begin{spacing}{2}
    \begin{algorithm}[t!]
	\caption{Nonparametric Bootstrap for Simultaneous Confidence Bands of $\hat{\beta}_{p}(s_{r}, t_{l})$} 
    {\fontsize{10}{15}\selectfont
	\begin{algorithmic}[1]
		\FOR {$b=1,\dots,B$}
            \STATE Resample the subject indexes from the index set $\{1, \dots, N\}$ with replacement and define $I_{b}$ be the set of indices;
            \STATE Denote the $b$th bootstrap sample as $\{Y_{I_{b}}(s_{r},t_{l}), \mathbf{X}_{I_{b}}(s_{r},t_{l})\}$;
            \STATE Use Step I and Step II of the estimation procedure to obtain $\hat{\beta}_{p}^{(b)}(s_{r},t_{l})$; 
            \ENDFOR
            \STATE Perform FPCA and marginal decomposition technique on $\{\hat{\beta}_{p}^{(1)}(s_{r},t_{l}), \dots, \hat{\beta}_{p}^{(B)}(s_{r},t_{l})\}$ to obtain $\{\hat{\psi}_{j}(s_{r}), \hat{\mathbf{b}}_{b,j}, j=1,\dots,J_{B}; \}$. Derive the mean function
            $\Bar{\beta}_{p}(s_{r},t_{l})$ and obtain the sample covariance of $\hat{\mathbf{b}}_{b,j}$, denoted by $\Bar{\mathbf{V}}_{j}$;
		        \FOR {$m=1,\ldots,M$}
			\STATE Generate a random variable $\mathbf{u}_{m,j}$ from the multivariate normal with mean $\boldsymbol{0}$ and covariance matrix $\Bar{\mathbf{V}}_{j}$;
				\STATE Derive $\hat{\beta}_{m,p}(s_{r},t_{l})=\Bar{\beta}_{p}(s_{r},t_{l}) +\sum_{j=1}^{J_{B}}\mathbf{B}^{T}_{j}(t_{l})\mathbf{u}_{m,j}\hat{\psi}_{j}(s_{r})$;
                    \STATE Compute 
                    $q^{\ast}_{m} = \max_{s_{r},t_{l}}\{|\hat{\beta}_{m,p}(s_{r},t_{l}) - \hat{\beta}_{p}(s_{r},t_{l})|/\widehat{\text{Var}}\{\hat{\beta}_{p}(s_{r},t_{l})\}^{1/2}\}$;
                \ENDFOR
            \STATE Obtain the $100(1 - \alpha)\%$ empirical quantile of $\{q^{\ast}_{1}, \dots, q^{\ast}_{M}\}$, denoted by $q_{1-\alpha}$;
            \STATE The $100(1 - \alpha)\%$ simultaneous confidence bands are given by $\hat{\beta}_{p}(s_{r},t_{l}) \pm q_{1-\alpha}\widehat{\text{Var}}\{\hat{\beta}_{p}(s_{r},t_{l})\}^{1/2}$. 
	\end{algorithmic}
    }
    \label{alg:scb}
\end{algorithm}
\end{spacing}

To address the issue, we refer to simultaneous confidence bands (SCB) to account for correlation, which are commonly constructed using nonparametric bootstrap approaches. One method involves multivariate normal simulations (\citealp{crainiceanu2012}), with high computational cost due to the dimensionality equalling the sampling density of the functional domain. We use parameter simulations based on the number of B-splines basis functions and reduce to a tractable number of dimensions (\citealp{park2018, cui, crainiceanu2024}). The details of the bootstrap algorithm are provided in the Algorithm \ref{alg:scb}. Our bootstrap SCB algorithm makes construction computationally practical by leveraging B-splines technique. In addition, it can also serve as a data-driven inference tool for formal global tests about the coefficient functions without relying on distributional assumptions (\citealp{park2018,sergazinov2023}).

\subsection{Asymptotic Results}

In this section, we derive the asymptotic distribution of our pointwise-smoothing estimator by showing the asymptotic bias and variance structure. Asymptotics of our estimator is established based on the properties of the least square estimator and sandwich smoother which is equivalent to a bivariate kernel regression estimator with a product kernel,  $(RLh_{R}h_{L})^{-1}\sum_{r,l}\beta_{p}(s_{r},t_{l})H_{m_{R}}\{h^{-1}_{R}(s-s_{r})\}H_{m_{L}}\{h^{-1}_{L}(t-t_{l})\}$, where $H_{m}$ is the equivalent kernel for univariate penalized splines, $h_{R}$ and $h_{L}$ are the bandwidths (\citealp{wang2011, xiao}). The kernel function $H_{m}$ is symmetric and bounded. For simplicity, our results are for the case of equally spaced design points and knots. For notation convenience, $a \sim b$ means $a/b$ converges to 1. 

We first derive the asymptotic bias in the interior points. Let $m_{R}$ and $m_{L}$ are difference orders of differencing matrices, $m_{T} = 4m_{R}m_{L} + m_{R} + m_{L}$ for notation simplicity.

\textit{Propostion 1. Suppose conditions (a)-(d) and (g) are satisfied, further assume $K_{R} \sim c_{R}(RL)^{b_{1}}$, $K_{L} \sim c_{L}(RL)^{b_{2}}$, with $b_{1} > (m_{R} + 1)m_{L}/m_{T}$, $b_{2} > (m_{L} + 1)m_{R}/m_{T}$, $h_{R} \sim d_{R}(RL)^{-m_{L}/m_{T}}$, $h_{L} \sim d_{L}(RL)^{-m_{R}/m_{T}}$ for some positive constants $c_{R}, c_{L}, d_{R}, d_{L}$. Then, for any $(s,t) \in (0,1)^{2}$, we have 
    \begin{equation*}
    \begin{aligned}
          \textnormal{bias}\{\hat{\beta}_{p}(s,t)\} & = (-1)^{m_{R} + 1}d^{2m_{R}}_{R}\frac{\partial^{2m_{R}}}{\partial s^{2m_{R}}}\beta_{p}(s,t) + (-1)^{m_{L} + 1}d^{2m_{L}}_{L}\frac{\partial^{2m_{L}}}{\partial t^{2m_{L}}}\beta_{p}(s,t) + o(h^{2m_{L}}_{L}).
    \end{aligned}
    \end{equation*}
}The bias remians same with the sandwich smoother (\citealp{xiao}), containing componests from $\mathcal{S}$ and $\mathcal{T}$. Noted that the bias converges at a slower rate at the boundary than in the interior (\citealp{li2008}), the proof of which is ignored here. 

To derive the asymptotic variance of the estimator, we assume the covariates are identically and independently distributed as well as time-invariant. For simplicity of illustration, we also assume no missing values. 

\textit{Proposition 2. Suppose conditions (a)-(g) are satisfied and under the conditions of Proposition 1, when $N \rightarrow \infty$, we have
\begin{equation*}
      \textnormal{var}\{\hat{\beta}_{p}(s,t)\} =  2(RLh_{R}h_{L}N)^{-1}\omega_{p}\sigma^{2}(s,t)\kappa(H_{m_{R}})\kappa(H_{m_{L}}) + o(h_{L}^{4m_{L}}).
\end{equation*}
}where $\omega_{p}$ is the $(p,p)$th entry of $\Omega = \mathbb{E}(\mathbf{X}_{1,s_{r},t_{l}}\mathbf{X}_{1,s_{r},t_{l}}^{T})$ and $\kappa(H_{m}) = \int H^{2}_{m}(u)du$. The proposition implies that the asymptotic variance structure of our estimator has an extra component because of the existence of pointwise least square estimator, compared to the sandwich smoother only (\citealp{xiao}). Additionally, it also shows that the correlation influence of two points can be ignored, similarly with kernel regression estimator (\citealp{wand1995}). 

Based on the asymptotic bias and variance structures, and proposition 1 of \cite{xiao}, the corresponding asymptotic distribution of our estimator is given by 
\begin{equation*}
    (RL)^{2m_{R}m_{L}/m_{T}}(\hat{\beta}_{p}(s,t) - \beta_{p}(s,t)) \rightarrow N(\alpha_{p}(s,t), V_{p}(s,t)), 
\end{equation*}
in distribution as $R \rightarrow \infty$ and $L \rightarrow \infty$, $N \rightarrow \infty$, where $\alpha_{p}(s,t) = (-1)^{m_{R} + 1}d^{2m_{R}}_{R}\frac{\partial^{2m_{R}}}{\partial s^{2m_{R}}}\beta_{p}(s,t) + (-1)^{m_{L} + 1}d^{2m_{L}}_{L}\frac{\partial^{2m_{L}}}{\partial t^{2m_{L}}}\beta_{p}(s,t)$ and $V_{p}(s,t) = 2(RLh_{R}h_{L}N)^{-1}\omega_{p}\sigma^{2}(s,t)\kappa(H_{m_{R}})\kappa(H_{m_{L}})$.

\section{Simulation}
\label{sec:simu}

We conduct extensive simulation studies to evaluate the performance of the proposed estimation and inference procedure. Our method is examined not only in bivariate but also univariate perspectives, as other competing FMEM estimation methods often only consider fixed effects over functional domain.

The bivariate functional model is simulated as follows:
\begin{equation*}
    Y_{i}(s,t) = \beta_{0}(s,t) + X_{i}(s)\beta_{1}(s,t) + \gamma_{i0}(t) + z_{i}(s)\gamma_{i1}(t) + \epsilon_{i}(s, t), \ \ (s, t) \in [0, 1]^{2}.
\end{equation*}
The fixed-effect covariates are generated from $X_{i}(s) \sim N(0, 4)$ and $z_{i}(s) = 6(s-0.5)^{2} + N(0,\rho^{2})$, where $\rho$ represents how noisy the signal of repeatedly measured visits is. For bivariate coefficient functions, we take account of two different types as shown in Figure \ref{fig:true_beta} of the supplementary material. The first scenario (S1) presents a continuous non-differentiable bivariate function with local zero regions (sparse), while the second scenario (S2) presents a smooth bivariate function. The random effects are simulated as $\gamma_{il}(t) = a_{i1}\phi_{1l}(t) + a_{i2}\phi_{2l}(t)$, $l=0,1$. We use the scaled orthonormal functions
\[
\begin{bmatrix}
\phi_{1}(t)\\
\phi_{2}(t)
\end{bmatrix}
\propto
\begin{cases}
\begin{bmatrix}
1.5 - \sin(2\pi t) - \cos(2\pi t) & \sin(4\pi t)
\end{bmatrix}^{T} & \text{if } l=0 \\
\begin{bmatrix}
\cos(2\pi t) & \sin(2\pi t) 
\end{bmatrix}^{T}  & \text{if } l=1
\end{cases}
\]
to capture the subject-level fluctuations. The random coefficients are generated from $a_{i1} \sim N(0, 2\sigma^{2}_{B})$ and $a_{i2} \sim N(0, \sigma^{2}_{B})$ respectively and $\sigma^{2}_{B}$ depends on the relative importance of random effect $\text{SNR}_{B}$. The measurement error $\epsilon_{ij}(s,t) \sim N(0, \sigma_{\epsilon}^{2})$, where $\sigma_{\epsilon}^{2}$ depends on signal-to-noise ratio $\text{SNR}_{\epsilon}$. Here $\text{SNR}_{B}$ is defined as the ratio of the standard deviation of fixed-effect and random-effect surfaces, while $\text{SNR}_{\epsilon}$ is the ratio of the standard deviation of all liner predictors and that of the measurement errors (\citealp{scheipl,cui}). We set $\text{SNR}_{B} = \text{SNR}_{\epsilon} = 1$.

The performance of the method is evaluated from three aspects reflecting the accuracy of estimation and inference, as well as computational efficiency. First, the estimation error is assessed by integrated squared error (ISE) to measure the difference between the estimate and the underlying truth, defined as $\text{ISE}(\hat{\beta}_{p}) = |\mathcal{S}\times\mathcal{T}|^{-1}\int_{\mathcal{S}}\int_{\mathcal{T}}\big(\hat{\beta}_{p}(s,t) - \beta_{p}(s,t)\big)^{2}dt$, where $|\mathcal{S} \times \mathcal{T}|$ denotes the area of the entire domain. Secondly, the proportion of pointwise surfaces wrapping the true plane in the sandwich form is computed for bivariate functional slope to evaluate inferential performance on fine grids. We use the empirical coverage probability of 95\% PCB, defined by , defined by $|\mathcal{S}\times\mathcal{T}|^{-1}\int_{\mathcal{S}}\int_{\mathcal{T}}\mathbbm{1}_{\hat{\beta}_{p}(s,t) \in \text{PCB}_{p}(s,t)}dsdt$. Additionally, to measure the width of confidence bands we also report integrated actual width (IAW), defined as $|\mathcal{S}\times\mathcal{T}|^{-1}\int_{\mathcal{S}}\int_{\mathcal{T}} \{\widehat{\text{UB}}_{p}(s,t) - \widehat{\text{LB}}_{p}(s,t)\}dsdt$, where $\widehat{\text{UB}}_{p}(\cdot,\cdot)$ and $\widehat{\text{LB}}_{p}(\cdot,\cdot)$ are pointwise estimates of upper bound and lower bound respectively. When comparing with other methods for FMEM from the univariate perspective, we accommodate the above metrics with only $\mathcal{T}$ direction. Computing time for the entire estimation procedure will also be counted to present the computational cost. 

\subsection{Bivariate Comparative Analysis}

We compare our proposed approach (2dFMM) with the concurrent bivariate functional regression method denoted by 2dGAM, which is developed by tensor product smooths (\citealp{ivanescu}). The following several simulation scenarios are considered. We let sample size $N \in \{50, 75, 100\}$, the number of functional grids $L \in \{100, 150, 200\}$, and the number of longitudinal grids $R \in \{10, 15, 20\}$. The baseline setting is $N = 50$, $R = 10$, and $L=100$, where all other sample generating parameters are fixed at their baseline values when one is changed. The noise argument of the longitudinal signal is set $\rho = 0.5$. A total of 100 replicates are independently simulated. 

\begin{figure}[!t]
\centering \hspace*{-0.5cm}
    \includegraphics[width=1.05\textwidth, height=0.2625\textwidth]{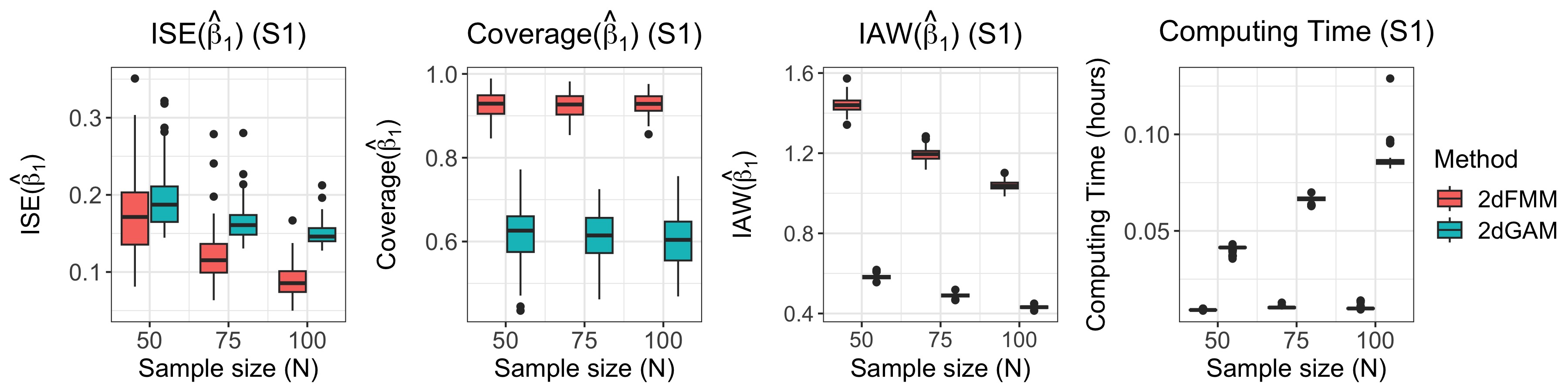}
\centering \hspace*{-0.5cm}
    \includegraphics[width=1.05\textwidth, height=0.2625\textwidth]{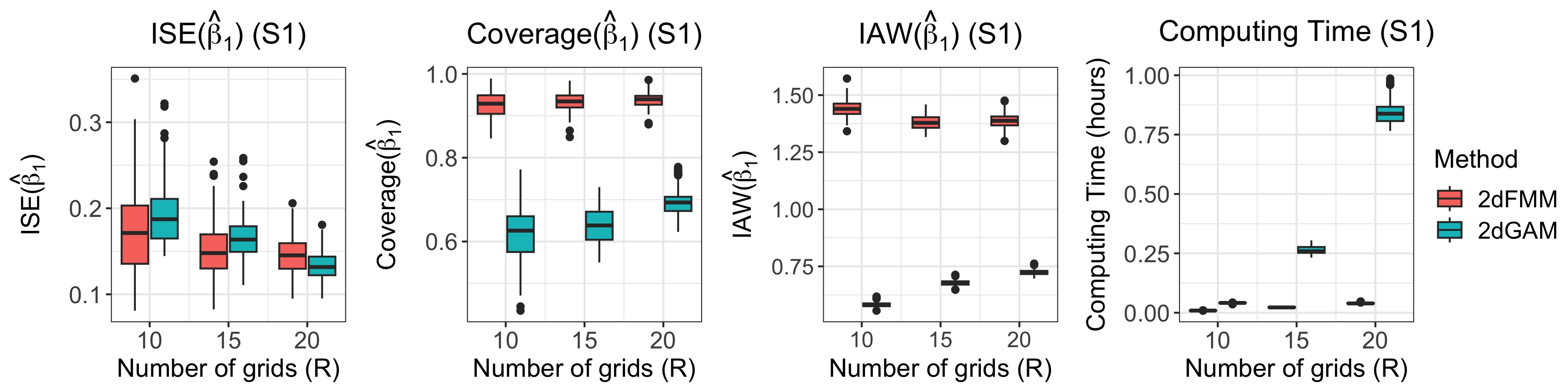}
\centering \hspace*{-0.5cm}
    \includegraphics[width=1.05\textwidth, height=0.2625\textwidth]{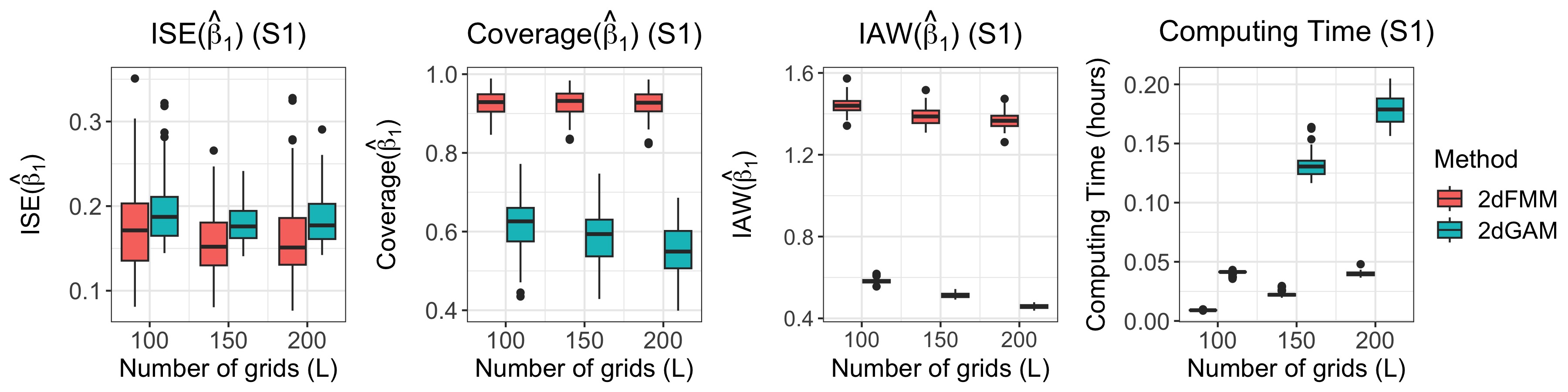}
\caption{The comparison of ISE, coverage probability of 95\% PCB (Coverage), IAW, and computing time for 2dFMM and 2dGAM under the first scenario (S1) of $\beta_{1}(s,t)$. The baseline setting is $N = 50$, $R = 10$, and $L=100$. When one parameter is changed, all other sample generating parameters are fixed at their baseline values. }
\label{fig:simu1S1}
\end{figure}

Figure \ref{fig:simu1S1} shows better performance of the proposed 2dFMM compared with 2dGAM regarding estimation accuracy in most of the cases under the first scenario. It is primarily attributed to the choice of sandwich smoother depending on differencing matrices, which can tackle this non-differentiable function. When the difference of dimensions between longitudinal and functional domains reduces, 2dGAM improves and slightly outperforms our method because the technique of tensor product spline basis expansions is appropriate for symmetric grids. The coverage probability of 95\% PCB for 2dFMM approaches the nominal level, while 2dGAM is far below it due to the ignorance of the four-dimensional correlation structure, leading to the too-narrow width of confidence bands. Since the 2dGAM model estimation depends on a representation of the generalized additive model, the computing time is not surprisingly much longer. The storage of intermediate large matrices is also problematic as it easily runs out of memory even when the sample generating parameters are not very large.

\begin{figure}[!t]
\centering \hspace*{-0.5cm}
    \includegraphics[width=1.05\textwidth, height=0.2625\textwidth]{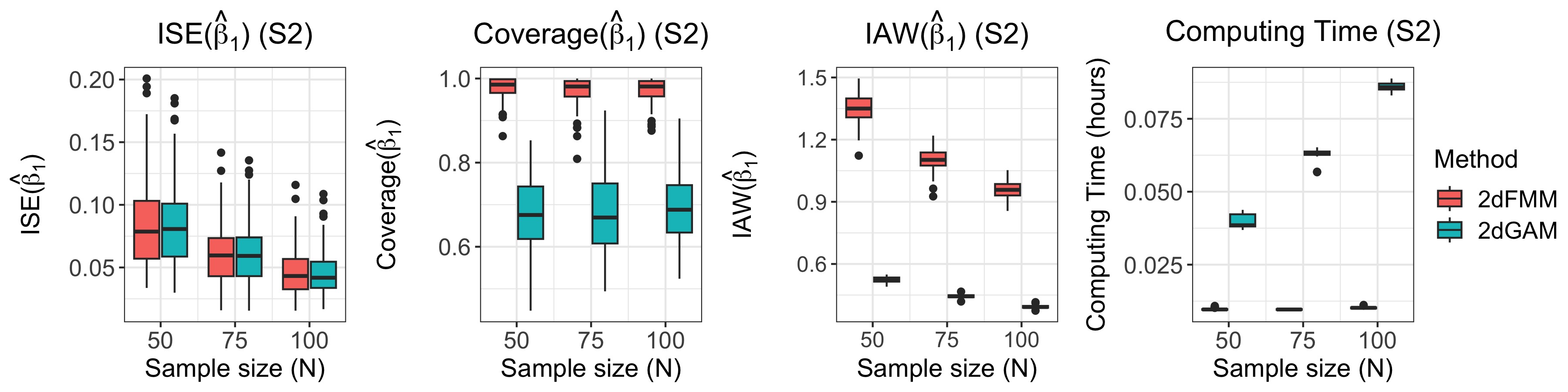}
\centering \hspace*{-0.5cm}
    \includegraphics[width=1.05\textwidth, height=0.2625\textwidth]{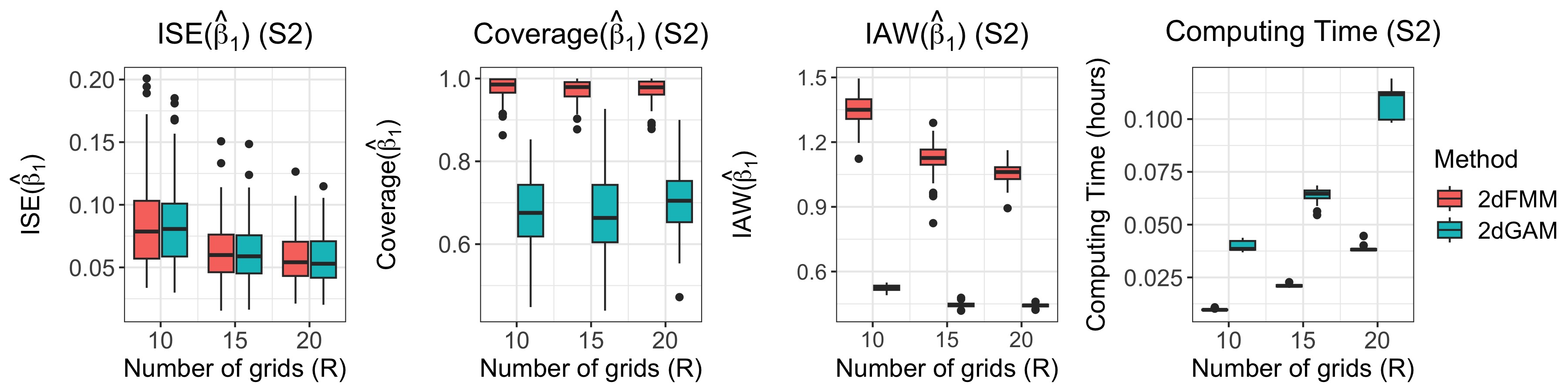}
\centering \hspace*{-0.5cm}
    \includegraphics[width=1.05\textwidth, height=0.2625\textwidth]{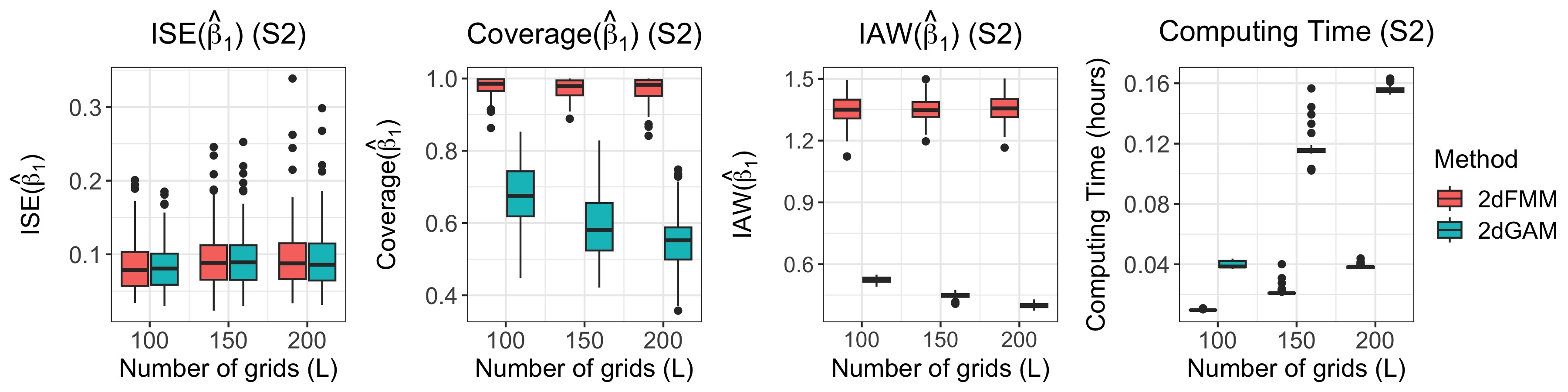}
\caption{The comparison of ISE, coverage probability of 95\% PCB (Coverage), IAW, and computing time for 2dFMM and 2dGAM under the second scenario (S2) of $\beta_{1}(s,t)$. The baseline setting is $N = 50$, $R = 10$, and $L=100$. When one parameter is changed, all other sample generating parameters are fixed at their baseline values. }
\label{fig:simu1S2}
\end{figure}

Figure \ref{fig:simu1S2} displays the results of the second scenario of the slope coefficient function, while we use tensor product smooths as post-smoother instead. The comparable performance of the two methods regarding estimation results indicates that the smoother is sufficient to compensate for the violation of the independence assumption underneath the pointwise technique when encountering continuous functions. The choice of tensor product smooths also solves the problem of symmetric numbers of longitudinal and functional grids that sandwich smoother has. Despite similar estimation accuracy, our method is still plausible given the nice inferential behaviors and low computational cost. 

In addition, we also perform the empirical coverage probability of 95\% pointwise and simultaneous confidence bands in different scenarios. The simultaneous confidence bands reach the nominal level.  The detailed comparisons between them are shown in Table \ref{table:simu_psb&scb} of the supplementary file.

\subsection{Univariate Comparative Analysis}

In longitudinal functional data analysis, fixed effects in FMEM framework are often evaluated only over the functional domain. Several methods including functional additive mixed models (FAMM), fast univariate inference (FUI), and fixed-effect inference for longitudinal functional data (FILF) allowing between-visit correlations are considered for comparisons, while FUI and FILF are designed for simpler computation (\citealp{scheipl, cui, li2022}). They are prespecified to incorporate random slope covariates using longitudinal time points. However, the performance of FAMM is not shown because it is similar to FUI while taking dramatically longer computing time and narrower confidence bands (\citealp{cui, li2022}). 

All approaches are evaluated in two cases: (i) bivariate functional slope is retained in the true model, but only the marginal effect over the functional domain is examined. (ii) bivariate functional intercept and slope are shrunk to univariate ones in the true model, i.e. $\beta_{p}(t) = |\mathcal{S}|^{-1}\int_{\mathcal{S}}\beta_{p}(s,t)ds, p=0,1$. Let the noise argument $\rho \in \{0.5, 2, 6\}$ control the magnitude of the longitudinal correlation in the true model. A total of 100 replicates are independently simulated.

\begin{figure}[!h]
\centering
\begin{subfigure}[b]{1\textwidth}
   \includegraphics[width=0.96\textwidth, height=0.32\textwidth]{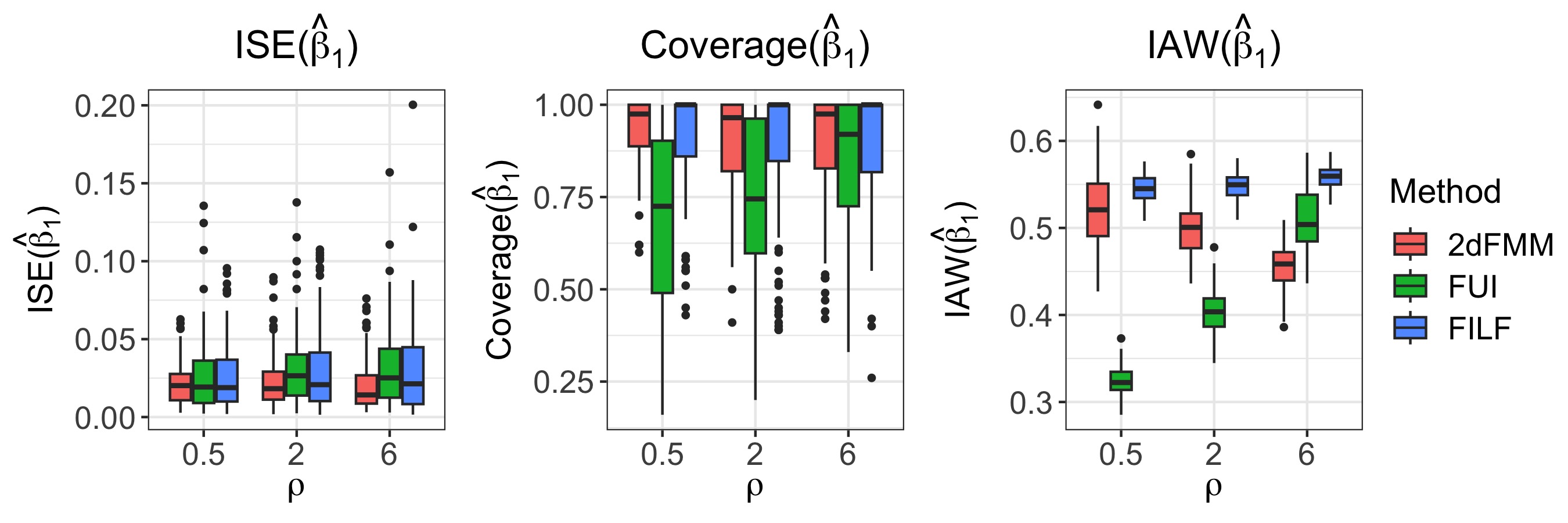}
   \caption{Case (i): the bivariate functional slope $\beta_{1}(s,t)$ under the second scenario (S2) is retained in the true model}
\end{subfigure}
\begin{subfigure}[b]{1\textwidth}
    \includegraphics[width=0.96\textwidth, height=0.32\textwidth]{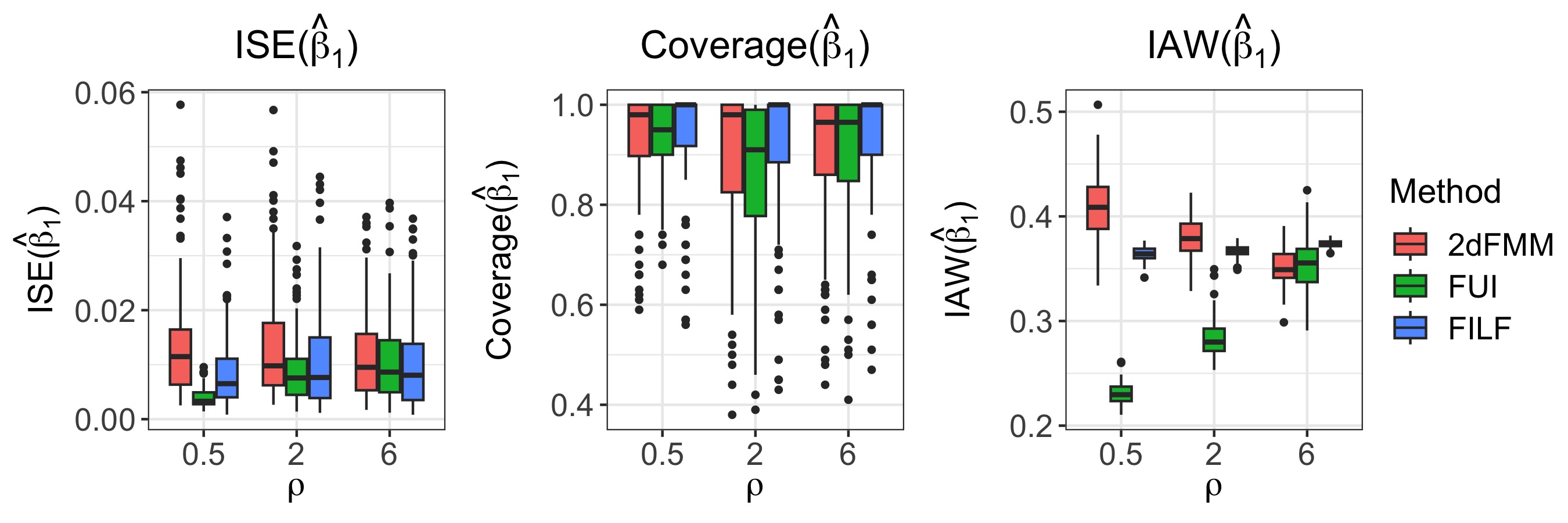}
   \caption{Case (ii): the bivariate functional slope $\beta_{1}(s,t)$ is shrunk to univariate $\beta_{1}(t)$ under the second scenario (S2) in the true model.}
\end{subfigure}
\caption{The comparisons of ISE, coverage probability of 95\% PCB  (Coverage), and IAW for FUI, FILF, and our method with reducing to functional direction with the setting $N = 50$, $R = 10$, and $L=100$. }
\label{fig:simu2_noise_S2}
\end{figure}

Figure \ref{fig:simu2_noise_S2}(a) presents the performance of different methods under case (i). It can be seen that the estimation accuracy of 2dFMM outperforms the other approaches at any strength of noise argument because of allowing two-dimensional effects. FMEM approaches are worse as the noise argument tends to be larger, indicating nearly no longitudinal correlation structures and therefore resulting in the misspecification of the functional random slope model. For confidence bands, our method is robust in terms of coverage level and actual width among all kinds of noise arguments, while FILF obtains high coverage level at the expense of wider confidence bands. On the other hand, Figure \ref{fig:simu2_noise_S2}(b) reveals the disadvantage of 2dFMM in case (ii) where a strong longitudinal correlation structure and univariate effect appear, which precisely meet the prespecified FMEM framework. The relatively low estimation accuracy is due to the misrecognition of the strong random longitudinal correlated signals as quantities of fixed effects. Given the reason, it is not surprising that the 2dFMM is better with heavier longitudinal noise, while others decline for the misspecification issue again. Despite the disadvantage, our method still presents decent coverage close to the nominal level in any case, even though it is slightly conservative in unfavorable conditions. The patterns remain unchanged for larger sample generating parameters (see Figure \ref{fig:simu2_noise_S2N100}-\ref{fig:simu2_noise_S2L200} of the supplementary material). 



\begin{table}[b!]
    \centering
    \scalebox{0.9}{
\begin{tabular}{c c c c c}
\hline
 &  & \multicolumn{3}{c}{Number of grids ($L$)} \\[0.5ex]
 \cline{3-5}
  Sample size ($N$) & Methods &  100 & 200 & 400 \\[0.5ex]
 \hline
100 & 2dFMM & 0.59 (0.01) & 2.40 (0.11) & 11.25 (0.49) \\[0.5ex]
 & 2dFMM (Parallel) & 0.55 (0.06) & 1.61 (0.07) & 7.49 (0.44) \\[0.5ex]
 & FUI  & 0.75 (0.02) & 2.72 (0.10) & 12.35 (1.01)\\[0.5ex]
 & FILF & 12.88 (2.39) & 13.52 (2.71) & 15.74 (0.86) \\[1.5ex]
200 & 2dFMM & 0.67 (0.02) & 2.58 (0.17) & 11.42 (0.32) \\[0.5ex]
 & 2dFMM (Parallel) & 0.57 (0.04) & 1.73 (0.18) & 7.54 (0.14) \\[0.5ex]
 & FUI  & 1.35 (0.02) & 5.01 (0.05) & $>$ 600 \\[0.5ex]
 & FILF & 97.51 (6.33) & 100.32 (12.64) & 119.31 (14.24) \\[1.5ex]
400 & 2dFMM & 1.12 (0.02) & 3.39 (0.36) & 12.48 (0.41) \\[0.5ex]
& 2dFMM (Parallel) & 0.88 (0.06) & 2.11 (0.07) & 8.18 (0.35)\\[0.5ex]
 & FUI  & 2.75 (0.03) &  11.45 (0.44) & $>$ 600 \\[0.5ex]
  & FILF &  &  &  \\[0.5ex]
 \hline
        \end{tabular}}
        \caption{Computing time (minutes) of 2dFMM and its parallel computing, FUI, and FILF in different scenarios of sample size ($N$) and the number of functional grids ($L$) under the second scenario (S2) of $\beta_{1}(s,t)$, given $R=10$ and $\rho=0.5$. Note that the value in the parenthesis corresponds to the standard deviation among 10 simulation replicates. Blank corresponds to out-of-computational memory. All simulation studies are carried out by Windows 10 2.30 GHz quad-core Intel Core i7 and 8 GB RAM. } 
        \label{table:comparison_time}
\end{table}

Table \ref{table:comparison_time} shows the computing time of these three methods with varying sample sizes and numbers of functional grids. The advantage of 2dFMM facing high-dimensional functional data is attributed to the pointwise-smoothing technique and marginal decomposition of the covariance function, while parallel implementation can further accelerate the process. Remarkably, the computation cost of our approach is not sensitive to either parameter, which causes particular concern in the analysis of our motivating data. However, FUI slows down given a considerable number of functional grids when encountering at least a moderate sample size. It is due to the fact that the estimation procedure for x for each pair of functional time points is computationally challenging once having a larger sample size. In addition, FILF uses the generalized additive model to refine the initial estimator, resulting in the same storage and cost limitation as 2dGAM.

\section{Application}
\label{sec:app}

In this section, we apply our proposed method to two studies for illustration. The first study uses motivating accelerometer data to examine the intraday and interday dynamic associations between adolescents' physical activity and their demographic characteristics, family socioeconomic status, and physical and mental health assessments. The second study uses a public environmental dataset to assess the association between electricity demand and temperature.

To present the statistical inferences of bivariate coefficient functions, we define a new metric $\hat{\text{I}}_{p}(s,t) = \hat{\beta}_{p}(s,t)\mathbbm{1}_{\{\widehat{\text{LB}}_{p}(s,t) > 0 \ \text{or} \ \widehat{\text{UB}}_{p}(s,t) < 0\}}$ to quantify and interpret the dynamicity of the associations. Heatmaps of the significance evaluation $\hat{\text{I}}_{p}(s,t)$ will be displayed: white regions indicate no significant effects, red indicates significantly positive effects, and blue indicates significantly negative effects, while the darkness of the colors represents the magnitude of the effects.

\subsection{Application to Shanghai School Adolescent Physical Activity Data}

In the Shanghai school adolescent study, we collected each subject's demographic information including a binary indicator of gender, grade from 7th to 12th, annual family income level, and mother's education level. Annual family income contains 7 levels from ``$\leq 10k$" to ``$\geq 300k$", while mother’s education level includes 8 levels from ``not graduated from primary school" to ``at least master degree", both of which are treated as ordinal variables. Mental health screening questionnaires were also conducted during measuring periods. Several common self-report health measurements are included: Depression Anxiety Stress Scales (DASS), Anticipatory and Consummatory Interpersonal Pleasure Scale (ACIPS), Emotion Regulation Questionnaire (ERQ) including two metrics: cognitive reappraisal and expressive suppression. All health assessment results are numeric to present the subjective psychological conditions within different aspects. 

We first regress the subjects' profiles to demographic and socioeconomic covariates in the baseline model.   To avoid the collinearity of mental health outcomes, each covariate is added to the baseline model separately. The functional response consists of a 16,191 $\times$ 1,440 dimensional matrix, where each row corresponds to ``day of week'' and each column corresponds to ``time of day''.

\begin{figure}[!b]
\centering
    \includegraphics[width=1\textwidth, height=0.45\textwidth]{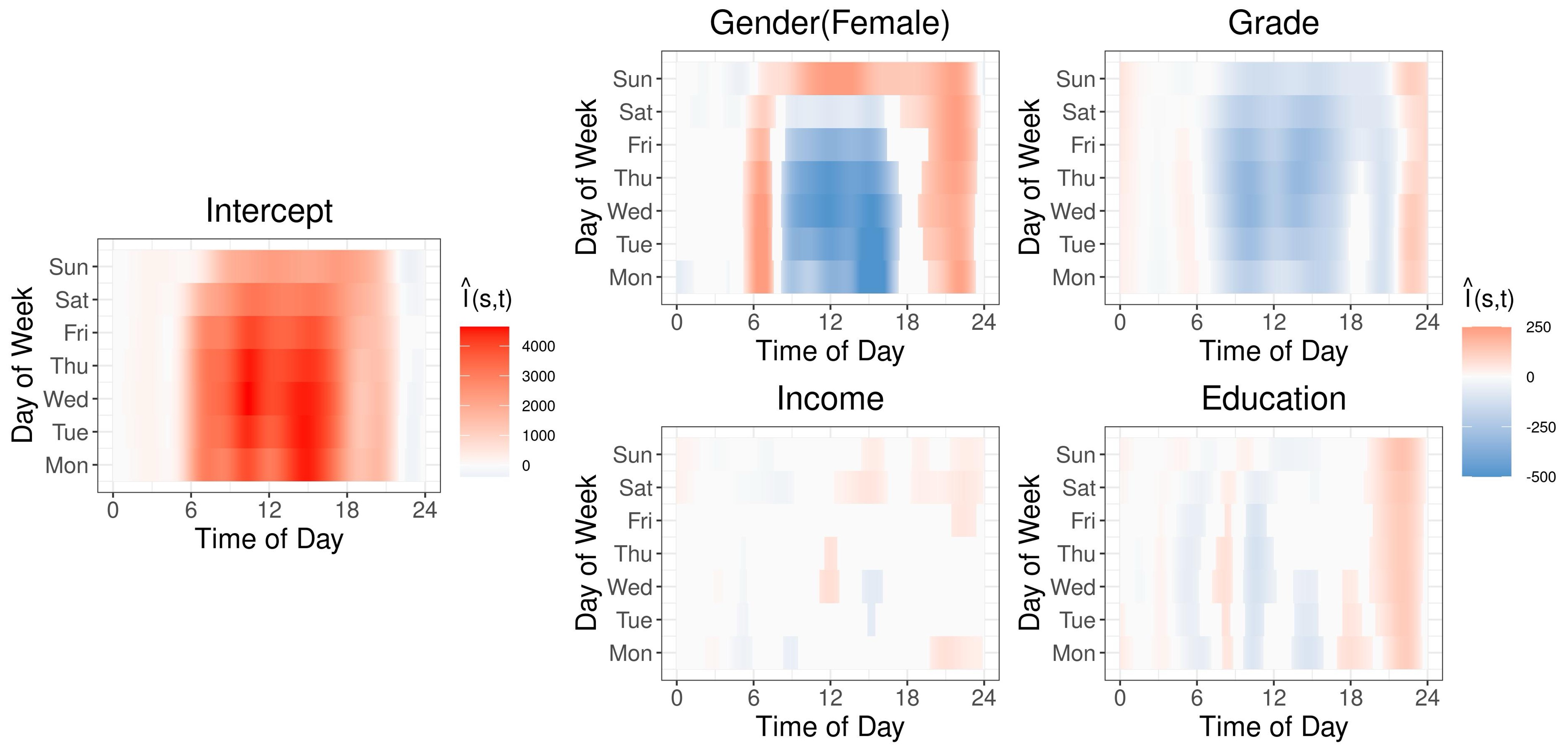}
    \caption{Statistical inference evaluation $\hat{\text{I}}_{p}(s,t)$ of intercept, demographic and socioeconomic covariates.}
\label{fig:real_baseline}
\end{figure}

\begin{figure}[!t]
\centering
    \includegraphics[width=1\textwidth, height=0.45\textwidth]{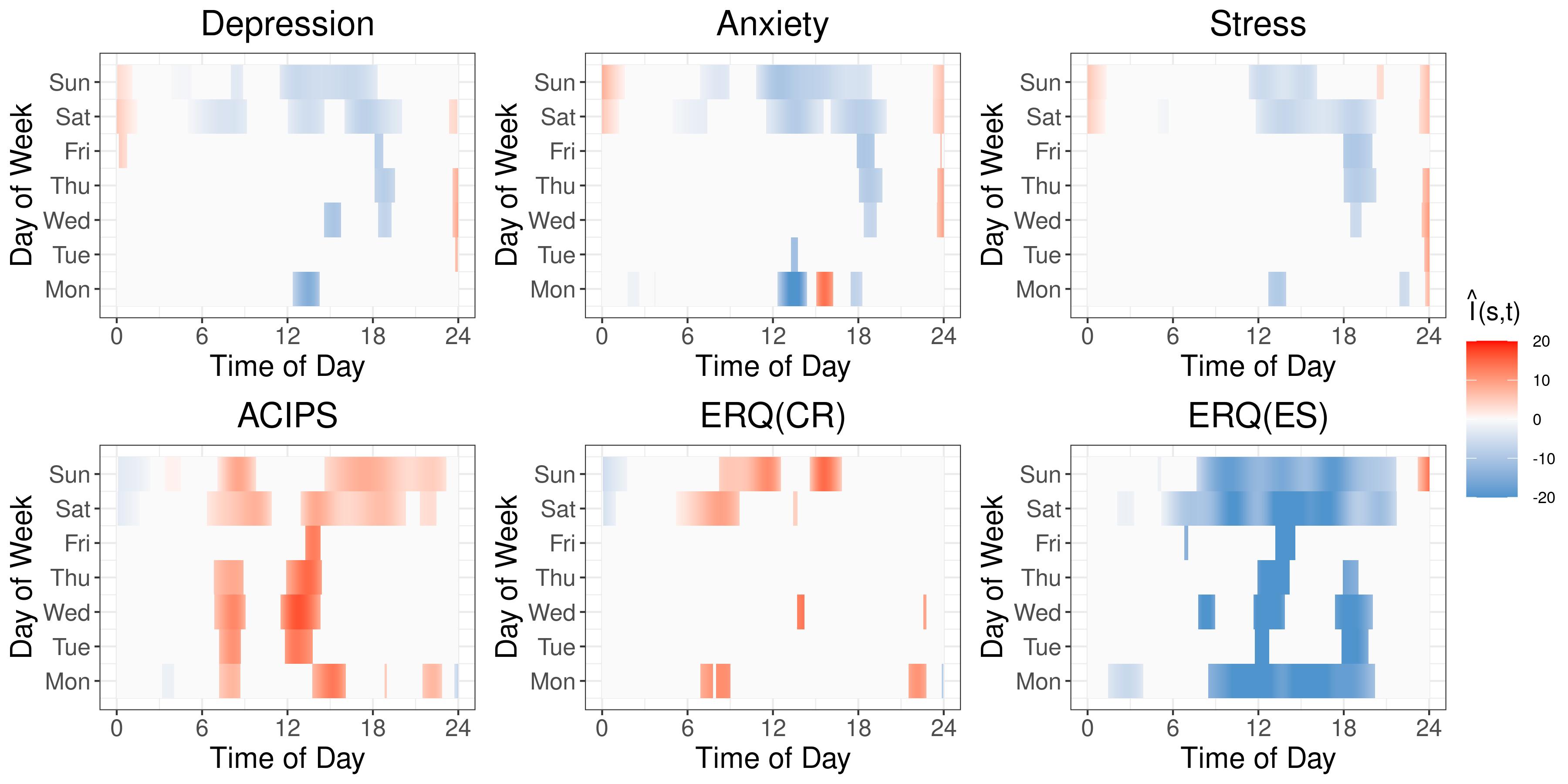}
    \caption{Statistical inference evaluation $\hat{\text{I}}_{p}(s,t)$ of mental health assessment results: DASS including Depression, Anxiety, and Stress, ACIPS (an interpersonal pleasure scale), ERQ scores of cognitive reappraisal (ERQ(CR)) and expressive suppression (ERQ(ES)).}
\label{fig:real_mental}
\end{figure}

Figure \ref{fig:real_baseline} shows the results of the baseline model, where only demographic data and socioeconomic status are incorporated. The evaluation heatmap of the estimated intercept function is consistent with the overall student physical activity patterns. It is observed that girls usually fall asleep later but wake up earlier, while boys are more vigorous in the daytime; yet it is surprising that girls essentially have more intense activity than boys on Saturday night and Sunday. Grade is an alternative characterization for age but has more substantial indications about school workloads and pressure. From the heatmap, elder students are critically more sedentary most of the time, except at 6 a.m., 6 p.m., and midnight, suggesting longer studying time due to laborious tasks. As for two socioeconomic covariates, the effect of family income level is weaker than mother's education level. As mohter's education level increases, students get up and go to bed later, probably because of more demanding assignments from family or richer material well-being such as commuting by private cars and more entertaining options at night. In addition, the results of the univariate effects along the time of day for only weekday or weekend in the baseline model are illustrated in Figure \ref{fig:real_baseline_weekday_weekend} of the supplementary material.


Figure \ref{fig:real_mental} demonstrates inferential heatmaps for self-reported health assessment results assessed by well-known questionnaires. The associations between physical activity and scores of subjective depression, anxiety, or stress (DASS) measures share similar patterns. Students tend to be more active at the weekend midnight if they have more severe symptoms of mental health. Nevertheless, being sedentary during 6 p.m. after school or on weekends is because a student who is considered symptomatic for DASS is likely reluctant to exercise. Anticipatory and Consummatory Interpersonal Pleasure Scale (AICPS) evaluates one's pleasant sensations. It is expected that students who own high ACIPS tend to be more active in the daytime, especially on weekends, revealing that happy students are typically more motivated and energetic. In addition, emotion regulation strategies (ERQ) measure respondents' propensity to adjust their emotions, comprising cognitive reappraisal (ERQ(CR)) and expressive suppression (ERQ(ES)). For example, heavily relying on expressive suppression leads to substantial physical activity reduction on weekends, while cognitive reappraisal mainly affects student behaviors during the weekend daytime. In addition, the results of the univariate effects for health assessment results along the time of day for only weekdays or weekends are illustrated in Figure \ref{fig:real_mental_weekday_weekend} of the supplementary material.

\subsection{Application to Electricity Demand Data}

In Adelaide, Australia, summer electricity demand exhibits high volatility and a strong correlation with temperature. Several studies have examined this relationship under various temperature conditions (\citealp{magnano2008, ivanescu}). The electricity demand and temperature data, accessible through the R \texttt{fds} package (\citealp{fds}), span from July 6, 1997, to March 31, 2007, providing half-hourly recordings for each day of the week. Our investigation focuses on whether electricity demand is associated with temperature and the impact of weekends as a binary indicator under our proposed modeling framework. The dataset comprises 63 two-dimensional samples, each representing a weekday of a year. The response variable is electricity demand, measured in megawatts, recorded at half-hourly intervals within each week of the year, resulting in 48 functional grids over a day and 52 longitudinal grids over a year. 

\begin{figure}[!t]
\centering
    \includegraphics[width=1\textwidth, height=0.238\textwidth]{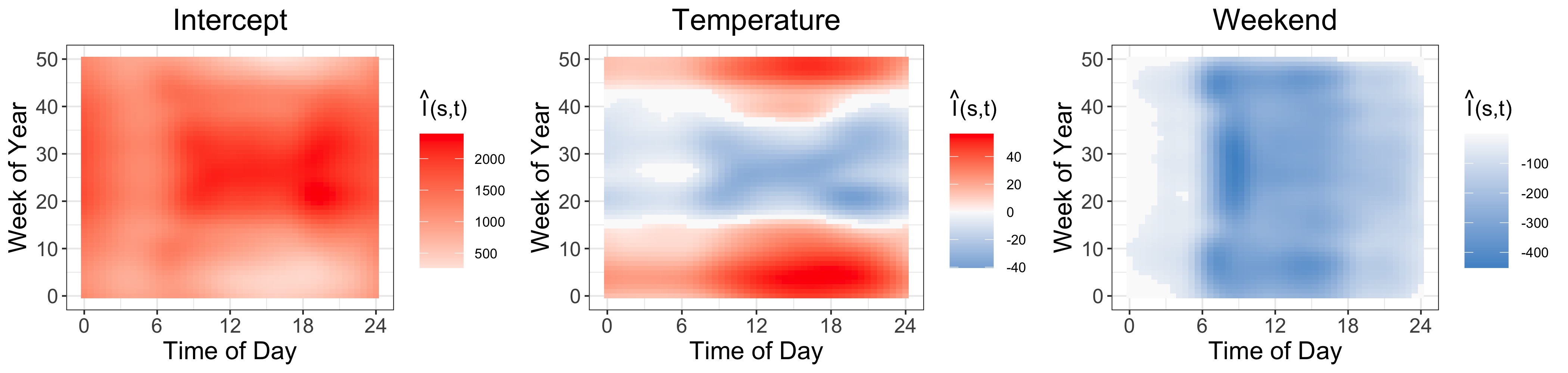}
    \caption{Statistical inference evaluation $\hat{\text{I}}_{p}(s,t)$ of intercept , temperature, and weekend.}
\label{fig:toy_app_bivariate}
\end{figure}

\begin{figure}[!b]
\centering
\begin{subfigure}[b]{1\textwidth}
   \includegraphics[width=1\textwidth, height=0.2775\textwidth]{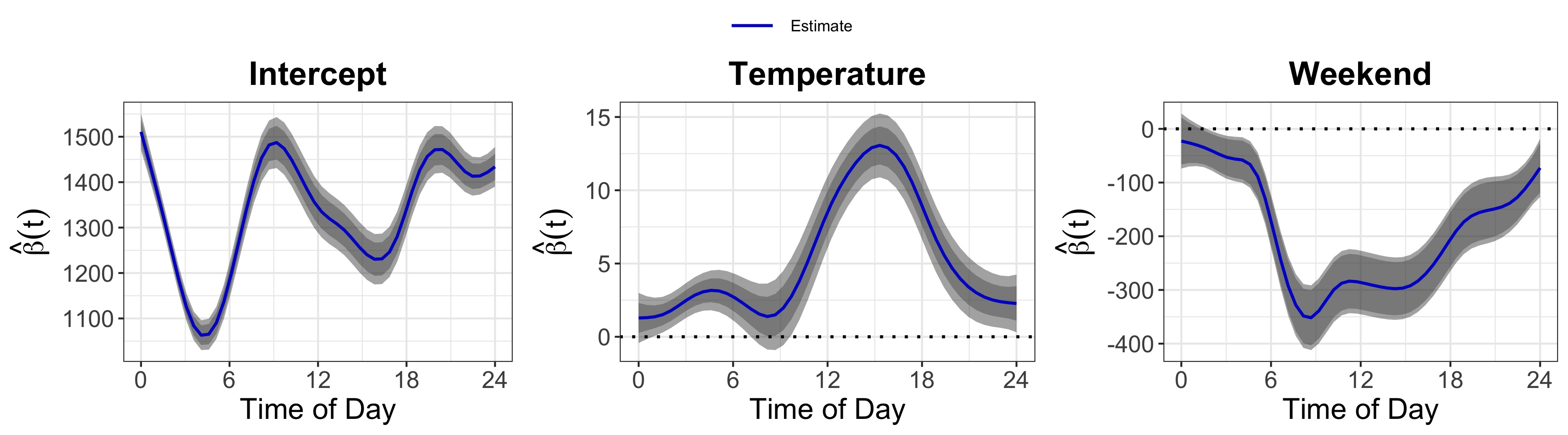}
   \caption{Unviarate effect over time of day.}
\end{subfigure}

\begin{subfigure}[b]{1\textwidth}
   \includegraphics[width=1\textwidth, height=0.2775\textwidth]{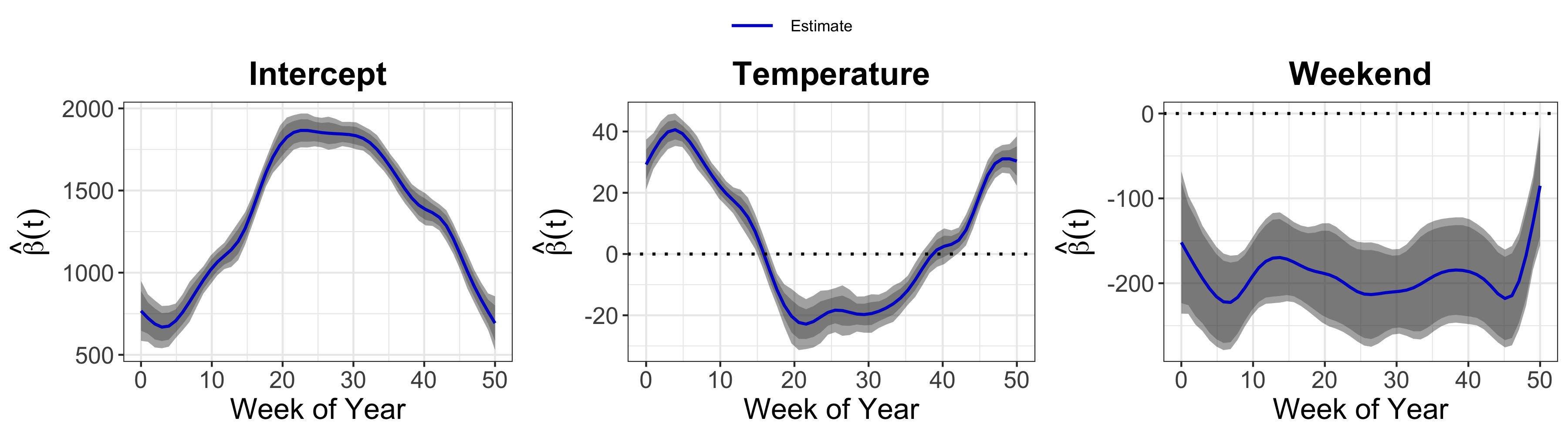}
   \caption{Unviarate effect over week of year.}
\end{subfigure}
\caption{Fixed-effects estimates (dashed blue line), 95\% PCB (dark gray shaded area), and 95\% SCB (light gray shaded area) of physical and intercept, temperature, and weekend.}

\label{fig:toy_app_uni}
\end{figure}

Figure \ref{fig:toy_app_bivariate} demonstrates the inferential heatmaps for intercept, temperature, and weekend. The intercept shows that electricity demand was generally higher during the mid-year weeks (Australian winter). During summer months, temperature had a strong positive effect on electricity demand, particularly from 10 a.m. to 8 p.m., suggesting a combination of factors including residential cooling needs and the population's daily activity patterns. Conversely, the mid-year weeks (Australian winter) show decreased demand with rising temperatures. This negative association may be attributed to Adelaide's pleasant winter temperatures, reducing the need for heating. Additionally, compared to weekdays, weekend corresponds to lower electricity demands throughout the entire year, especially in the morning around 6 a.m., which demonstrates the habit of waking up late on weekends. Figure \ref{fig:toy_app_uni} presents the univariate effects of time of day and week of year separately. While they provide strong supporting evidence for bivariate effects, the interplay of two temporal directions is overlooked. For example, the trend of prolonged positive effect was observed around week 10 and week 40 at 3 p.m. only from a bivariate perspective, possibly because work-related and other activities offset the climate effect.  

\section{Conclusion}
\label{sec:conc}

Motivated by daily activity profiles obtained by wearable device technologies, our work has provided an effective method for analyzing functional data over a series of times, which is common for longitudinal studies and known as repeatedly measured functional data. The analysis of this type of functional data is increasingly concerned in large-scale medical health research and many other studies such as biomedicine and environmental sciences. The development of efficient and flexible statistical tools for analyzing such data with ultra-high dimensionality and complex longitudinal-functional structure is an urgent problem. In this study, we aim to establish the two-dimensional functional mixed-effect model and efficient fixed-effect inference. We illustrate our method with the analysis of daily adolescent physical activity profiles and hourly electricity demands data, exploring their associations with various covariates of interests. 

The proposed fast three-stage estimation procedure sufficiently reduces computing costs for big data, especially with large-scale samples and functional grids. While we focus on the situation where the longitudinal sampling design is dense and regular in this study, real data is sometimes irregularly sampled. To address this, the estimation of bivariate coefficient functions can be adjusted by taking the average inside equal-size rectangular bins (\citealp{xiao}). Moreover, the covariance function can be estimated utilizing a local-linear smoothing approach and functional principal components analysis through conditional expectation (\citealp{yao2005}), while the remaining two-dimensional fixed-effect inference procedure is unchanged. Therefore, our method promotes the use of wearable devices in health research and has wider applications for longitudinal studies and spatial analysis.

\bigskip
\begin{center}
{\large\bf SUPPLEMENTARY MATERIAL}
\end{center}

\begin{description}

\item[Supplementary Material:] The Supplementary Materials contain results of additional simulation studies, additional analyses of Shanghai School Adolescent Physical Activity Data study, and the proof of the propositions.
\item[R-package for method and simulations:] All code for model implementation and simulation is available at \url{https://github.com/Cheng-0621/2dFMM}.

\end{description}

\bibliographystyle{agsm}
\bibliography{main}

\end{document}


\def\spacingset#1{\renewcommand{\baselinestretch}%
{#1}\small\normalsize} \spacingset{1}


\if1\blind
{
  \title{\bf Supplementary Material for ``An Efficient Two-Dimensional Functional Mixed-Effect Model Framework for repeatedly measured Functional Data''}
  \author{CHENG CAO \\
    School of Data Science, City University of Hong Kong \\
    and \\
    JIGUO CAO \\
    Department of Statistics and Actuarial Science, Simon Fraser University \\
    and \\
    HAO PAN,  YUNTING ZHANG, FAN JIANG \\
    Department of Developmental and Behavioral Pediatrics, \\ Shanghai Children’s Medical Center, \\School of Medicine, Shanghai Jiao Tong University \\ 
    and \\
    XINYUE LI \\
    School of Data Science, City University of Hong Kong}
  \maketitle
} \fi

\if0\blind
{
  \bigskip
  \bigskip
  \bigskip
  \title{\bf An Efficient Two-Dimensional Functional Mixed-Effect Model Framework for repeatedly measured Functional Data}
  \date{}
  \maketitle
  \medskip
} \fi

\bigskip

\spacingset{1.9} 
\section{Derivation of Equation (2)}
The four-dimensional covariance function is derived as follows. 
\begin{equation*}
\begin{aligned}
    C(s,t;u,v) &= \mathbb{E}[(Y_{i}(s,t) - \mu(s,t, \mathbf{X}_{i}))(Y_{i}(u,v) - \mu(u,v, \mathbf{X}_{i}))] \\
    & = \mathbb{E}[\sum_{j=1}^{\infty}\xi_{j}(t)\psi_{j}(s)\sum_{h}^{\infty}\xi_{h}(v)\psi_{h}(u)] \\ 
    & = \sum_{j=1}^{\infty}\sum_{h=1}^{\infty}\psi_{j}(s)\psi_{h}(u)\mathbb{E}[\xi_{j}(t)\xi_{h}(v)] \\
    &= \sum_{j=1}^{\infty}\psi_{j}(s)\psi_{j}(u)\mathbb{E}[\xi_{j}(t)\xi_{j}(v)] \\
    &= \sum_{j=1}^{\infty}\psi_{j}(s)\psi_{j}(u)\Theta_{j}(t,v).
\end{aligned}
\end{equation*}

\section{Proof of Propositions}



To demonstrate the asymptotic behaviors, the following conditions are required:
\vspace{12pt}
{

\noindent \textit{(a) There exists a constant $\delta > 0$ such that $\sup_{s,t}\mathbb{E}(|\beta_{p}(s,t)|^{2+\delta}) < \infty$}. 

\noindent \textit{(b) The bivariate coefficient function $\beta_{p}(s,t)$ is $2m$ times continuously differentiable, where $m =\max(m_{R}, m_{L})$}.

\noindent \textit{(c) The noise variance $\sigma^{2}(s,t)$ is continuous}.

\noindent \textit{(d) The design points are equally spaced $(s_r, t_l) = ((r-1/2)/R, (l-1/2)/L)$}.

\noindent \textit{(e) The $P$-dimensional covariates for $i$ is  $\mathbf{X}_{i} = (1, x_{i1}(s_{r},t_{l}), \dots, x_{iP}(s_{r},t_{l}))^{T}, i=1,\dots,N$ are independently with $\mathbb{E}(\mathbf{X}_{i,s_{r}t_{l}}\mathbf{X}^{T}_{i,s_{r}t_{l}})$ positive definite for a fixed $s_{r} \in \mathcal{S}$ and $t_{l} \in \mathcal{T}$.} 

\noindent \textit{(f) The covariate is time-invariant and satisfies (\textit{e})}.

\noindent \textit{(g) We assume $R \sim cL$, $h_{R} = K^{-1}_{R}(\lambda_{R}K_{R}R^{-1})^{1/(2m_{R})}$, $h_{L} = K^{-1}_{L}(\lambda_{L}K_{L}L^{-1})^{1/(2m_{L})}$, $h_{R} = O(R^{-a_1})$, and $h_{L} = O(L^{-a_2})$ for some constants $0 < c, a_1, a_2 < 1$, where $\lambda_{R}$ and $\lambda_{L}$ are tuning parameters for smoother. Also, $(K_{R}h^{2}_{R})^{-1} = o(1)$ and $(K_{L}h^{2}_{L})^{-1} = o(1)$.} 

}
\vspace{12pt}
Conditions (a)-(d) are just some regularity conditions for the asymptotic results and weakest possible conditions. They are imposed for the convenience of the technical proofs. Condition (d) assumes interior points. Condition (e) tells that each subject's covariates are \textit{i.i.d.} for a fixed sampling point. Condition (f) usually holds for our repeatedly measured data sets. Condition (g) is the standard assumption of kernel function to construct the sandwich smoother. 

\noindent \textit{Proof of Proposition 1}

From the proof of Theorem 1 in \cite{xiao}, it can show given $\tilde{\beta}_{p}(s,t)$, 
\begin{equation*}
    \mathbb{E}\{\hat{\beta}_{p}(s,t)|\tilde{\beta}_{p}(s,t)\} = \int\int \tilde{\beta}_{p}(s-h_{R}u, t-h_{L}v)H_{m_{R}}(u)H_{m_{L}}(v)dudv +O(\zeta_1) + O(\zeta_2),
\end{equation*}
where $\zeta_1= \max\{(Rh_{R})^{-2}, (Lh_{L})^{-2}\}$ and $\zeta_2=\max\{(K_{R}h_{R})^{-2}, (K_{L}h_{L})^{-2}\}$.

By law of total expectation, 
\begin{equation}
\begin{aligned}
        \mathbb{E}\{\hat{\beta}_{p}(s,t)\} & =  \mathbb{E}\{\mathbb{E}\{\hat{\beta}_{p}(s,t)|\tilde{\beta}_{p}(s,t)\}\} \\
        & = \int\int \mathbb{E}\{\tilde{\beta}_{p}(s-h_{R}u, t-h_{L}v)\}H_{m_{R}}(u)H_{m_{L}}(v)dudv +O(\zeta_1) + O(\zeta_2) \\
        & = \int\int \beta_{p}(s-h_{R}u, t-h_{L}v)H_{m_{R}}(u)H_{m_{L}}(v)dudv + O(\zeta_1) + O(\zeta_2).
        \label{eq:cond_mean}
\end{aligned}
\end{equation}
The last equation holds because of unbiased least square estimator $\tilde{\beta}_{p}(s,t)$. 
Taking the Taylor expansion of $\beta_{p}(s-h_{R}u, t-h_{L}v)$ at $(s,t)$ until the $2m_{R}$th and $2m_{L}$th partial derivative with respect to $s$ and $t$ respectively, we can cancel out these integrals. Therefore, the asymptotic bias is obtained, 
\begin{equation*}
\begin{aligned}
    \mathbb{E}\{\hat{\beta}_{p}(s,t)\} - \beta_{p}(s,t) = & (-1)^{m_{R} + 1}h^{2m_{R}}_{R}\frac{\partial^{2m_{R}}}{\partial s^{2m_{R}}}\beta_{p}(s,t) + (-1)^{m_{L} + 1}h^{2m_{L}}_{L}\frac{\partial^{2m_{L}}}{\partial t^{2m_{L}}}\beta_{p}(s,t) + \\
    & o(h^{2m_{R}}_{R}) + o(h^{2m_{L}}_{L}) +O(\zeta_1) + O(\zeta_2).
\end{aligned}
\end{equation*}
We let $h_{R}$ and $h_{L}$

Under some conditions of bandwidth $h_{R}, h_{L}$ and number of knots $K_{R}, K_{L}$, we can have $h^{2m_{R}}_{R}$, $h^{2m_{L}}_{L}$, $\zeta_{1}$, and $\zeta_{2}$ which are of the same order. Similarly,  $h^{4m_{L}}_{L}$ and $(RLh_{R}h_{L})^{-1}$ are of the same order as well, with details shown in \citet{xiao}. 

\noindent \textit{Proof of Proposition 2}

By law of total variance, 
\begin{equation*}
        \text{var}\{\hat{\beta}_{p}(s,t)\} = \mathbb{E}\{\text{var}(\hat{\beta}_{p}(s,t)|\tilde{\beta}_{p}(s,t))\} + \text{var}\{\mathbb{E}(\hat{\beta}_{p}(s,t)|\tilde{\beta}_{p}(s,t))\}.
\end{equation*}
By \eqref{eq:cond_mean} and conditions (e)-(f), we set $\Omega = \mathbb{E}(\mathbf{X}_{1,s_{r}t_{l}}\mathbf{X}_{1,s_{r}t_{l}}^{T})$ and $N\Omega \overset{p}{\rightarrow} \mathbf{X}_{s_{r},t_{l}}^{T}\mathbf{X}_{s_{r},t_{l}}$. Hence, $(\mathbf{X}_{s_{r},t_{l}}^{T}\mathbf{X}_{s_{r},t_{l}})^{-1} = N^{-1}\Omega^{-1}\{1+o_{p}(1)\}$, by the law of large numbers, we deduce that 
\begin{equation*}
    \text{var}\{\tilde{\beta}_{p}(s_{r},t_{l})\} = \sigma^{2}(s_{r},t_{l})N^{-1}e^{T}_{p}\Omega^{-1}e_{p}\{1 + o_{p}(1)\} = \sigma^{2}(s_{r},t_{l})N^{-1}\omega_{p}\{1 + o_{p}(1)\},
\end{equation*}
where $\omega_{p}=e^{T}_{p}\Omega^{-1}e_{p}, p=0,1,\dots,P$ and $\Omega$ is positive definite, where $\omega_{p}$ is the $(p,p)$th entry of $\Omega^{-1}$. 

The first term on the right-hand side follows Theorem 1 in \cite{xiao},
\begin{equation*}
\begin{aligned}
    \mathbb{E}\{\text{var}(\hat{\beta}_{p}(s,t)|\tilde{\beta}_{p}(s,t))\} & = \mathbb{E}\{ \text{var}\{\tilde{\beta}_{p}(s_{r},t_{l})\}
    \kappa(H_{m_{R}})\kappa(H_{m_{L}}) + o((RLh_{R}h_{L})^{-1})\} \\ 
    & = \frac{\sigma^{2}(s_{r},t_{l})\omega_{p}}{NRLh_{R}h_{L}}\kappa(H_{m_{R}})\kappa(H_{m_{L}}) + o_{p}(N^{-1}h_{L}^{4m_{L}}) + o(h_{L}^{4m_{L}}),
\end{aligned}
\end{equation*}
where $\kappa(H_{m}) = \int H^{2}_{m}(u)du$. 

The second term of the right-hand side is
\begin{equation*}
\begin{aligned}
     \text{var}\{\mathbb{E}(\hat{\beta}_{p}(s,t)|\tilde{\beta}_{p}(s,t))\} & = \text{var}\{\frac{1}{RLh_{R}h_{L}}\sum_{r,l}\tilde{\beta}_{p}(s_{r},t_{l})H_{m_{R}}\big(\frac{s-s_r}{h_{R}}\big)H_{m_{L}}\big(\frac{t-t_l}{h_{L}}\big) + O(\zeta_1) + O(\zeta_2)\} \\ 
    & = \frac{1}{RLh_{R}h_{L}} \text{var}\{\tilde{\beta}_{p}(s_{r},t_{l})\} \kappa(H_{m_{R}})\kappa(H_{m_{L}}) + o((RLh_{R}h_{L})^{-1})\\ 
    & = \frac{\sigma^{2}(s,t)\omega_{p}}{RLh_{R}h_{L}N}\{1 + o_{p}(1)\} \kappa(H_{m_{R}})\kappa(H_{m_{L}}) + o((RLh_{R}h_{L})^{-1}) \\
    & =  \frac{\sigma^{2}(s,t)\omega_{p}}{RLh_{R}h_{L}N}\kappa(H_{m_{R}})\kappa(H_{m_{L}}) + o_{p}(N^{-1}h_{L}^{4m_{L}}) + o(h_{L}^{4m_{L}}).
\end{aligned}
\end{equation*}
The second equality holds by the derivations in \cite{wand1995}.

To sum up these two terms, as $N \rightarrow \infty$, we have
\begin{equation*}
      \text{var}\{\hat{\beta}_{p}(s,t)\} =  2(RLh_{R}h_{L}N)^{-1}\omega_{p}\sigma^{2}(s,t)\kappa(H_{m_{R}})\kappa(H_{m_{L}}) + o(h_{L}^{4m_{L}}).
\end{equation*}

\section{Additional Information for Simulation}
\label{sec:ai_simu}

\begin{figure}[!htb]
\centering \hspace*{-1.2cm} 
  \includegraphics[width=1.1\textwidth, height=0.41\textwidth]{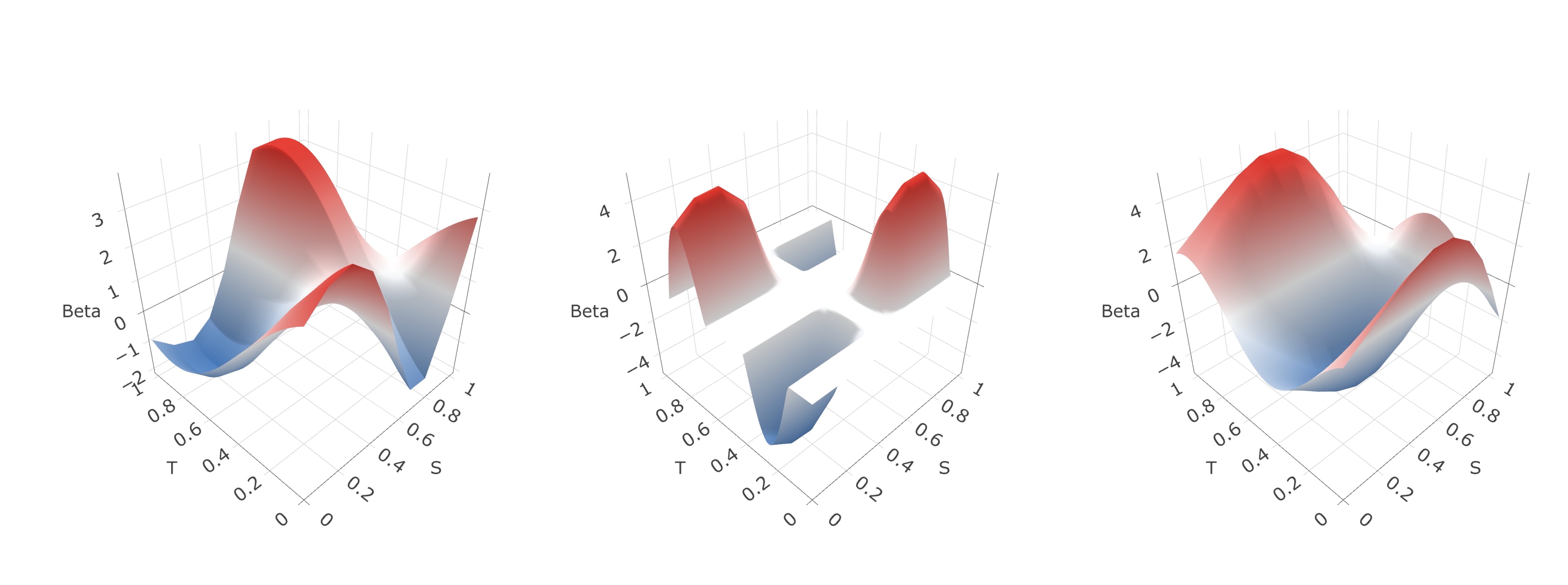}  
\caption{The left panel is the true bivariate functional intercept, and the middle and right panels are two types of true bivariate functional slope $\beta_{1}(s,t)$, denoted by S1 and S2. }
\label{fig:true_beta}
\end{figure}

Figure \ref{fig:true_beta} presents the true intercept surface and two different scenarios (S1 and S2) of slope surface, where S1 shows a continuous non-differentiable bivariate function with local zero regions and S2 shows a smooth bivariate function. For $(s, t) \in \mathcal{S} \times \mathcal{T} = [0,1]^{2}$, we define \\
(i) $\beta_{0}(s,t) = 3\sin(\pi(s+0.5)^{2})\cos(\pi t+0.5) + 1$; \\
(ii) S1: let $\alpha(s,t) = \sin(0.8\pi(s+0.5)^2)\cos(4\pi t)\mathbbm{1}_{\{(s, t) \in [0.2, 0.5] \times [0.14, 0.38]\}}$, then 
\[
\beta_{1}(s,t) 
\begin{cases}
5\alpha(s,t) & \text{for} \ (s, t) \in [0.1, 0.4] \times [0.14, 0.38] \cup (s, t) \in [0.7, 1] \times [0.62, 0.86] \\
-5\alpha(s,t) & \ \text{for} \ (s, t) \in [0.7, 1] \times [0.14, 0.38] \cup (s, t) \in [0.1, 0.4] \times [0.62, 0.86] \\
0 & \ \text{otherwise}  \\
\end{cases}
\]
has a zero subregion, indicating no impact on the response in this area; \\
(iii) S2: $\beta_{1}(s,t) = 5\sin(0.5\pi(s+0.5)^2)\cos(2\pi t+0.5)$ has no zero subregion but crossings at zero.

\begin{figure}[!thb]
\centering
\begin{subfigure}[b]{1\textwidth}
   \includegraphics[width=0.96\textwidth, height=0.32\textwidth]{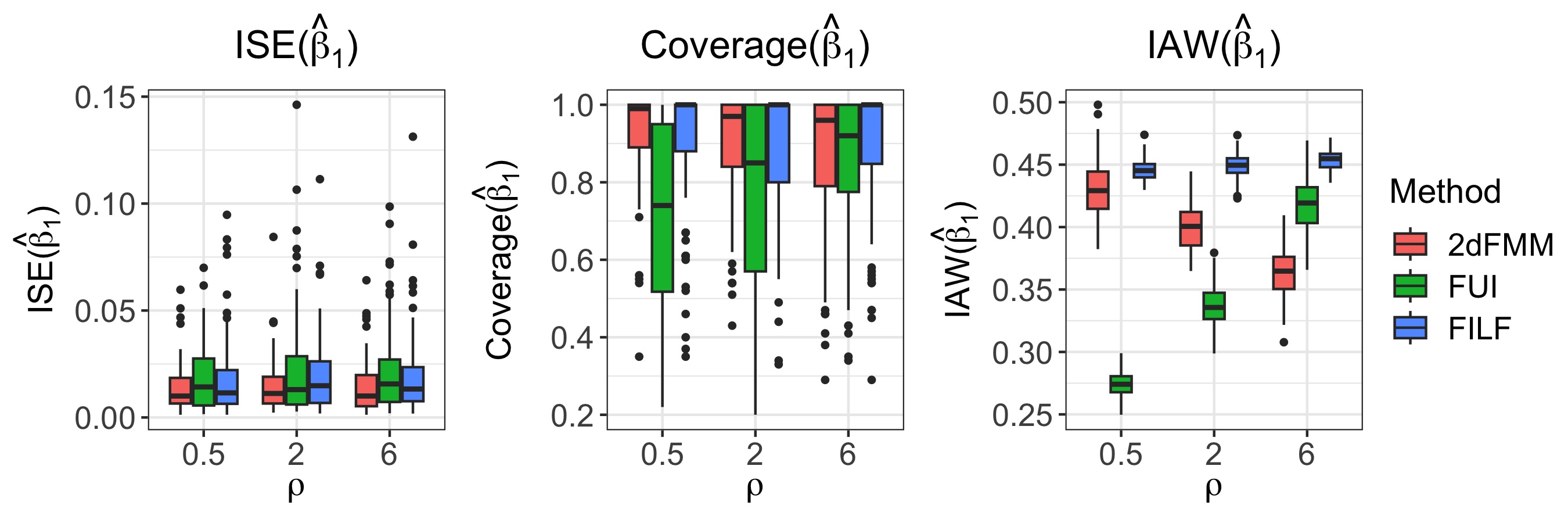}
   \caption{Case (i): the bivariate functional slope $\beta_{1}(s,t)$ under the second scenario (S2) is retained in the true model}
\end{subfigure}
\begin{subfigure}[b]{1\textwidth}
    \includegraphics[width=0.96\textwidth, height=0.32\textwidth]{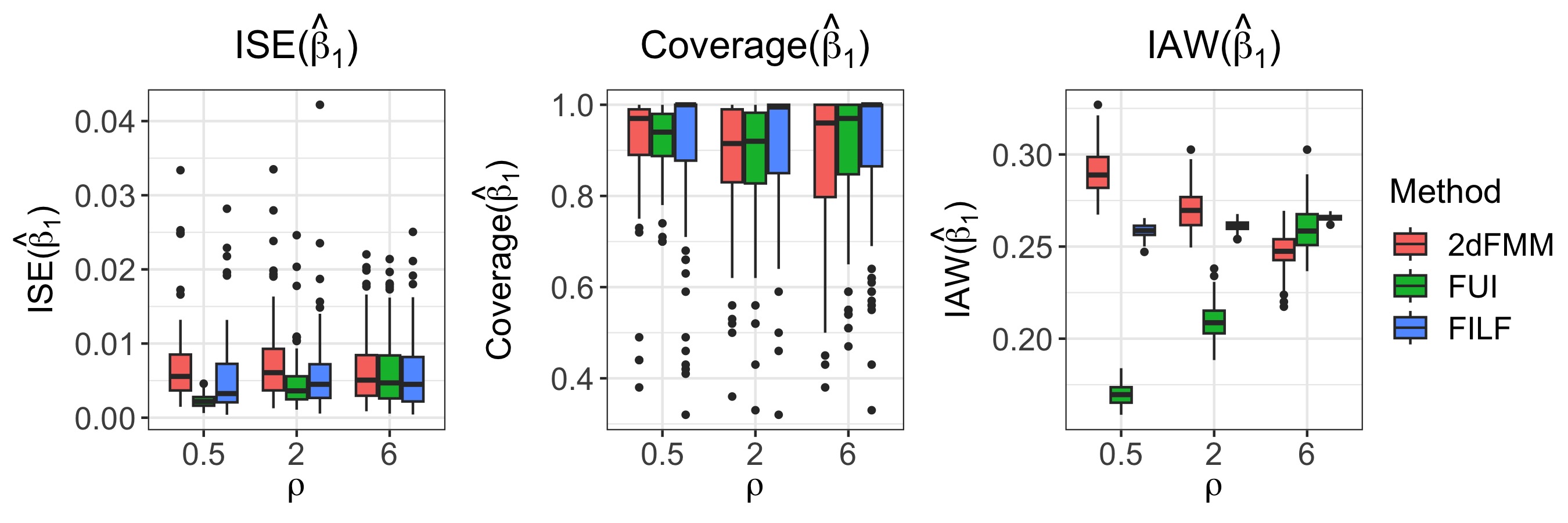}
   \caption{Case (ii): the bivariate functional slope $\beta_{1}(s,t)$ is shrunk to univariate $\beta_{1}(t)$ under the second scenario (S2) in the true model.}
\end{subfigure}
\caption{The comparisons of ISE, coverage probability of 95\% PCB  (Coverage), and IAW for FUI, FILF, and our method with reducing to functional direction with the setting $N = 100$, $R = 10$, and $L=100$. }
\label{fig:simu2_noise_S2N100}
\end{figure}

\begin{figure}[!thb]
\centering
\begin{subfigure}[b]{1\textwidth}
   \includegraphics[width=0.96\textwidth, height=0.32\textwidth]{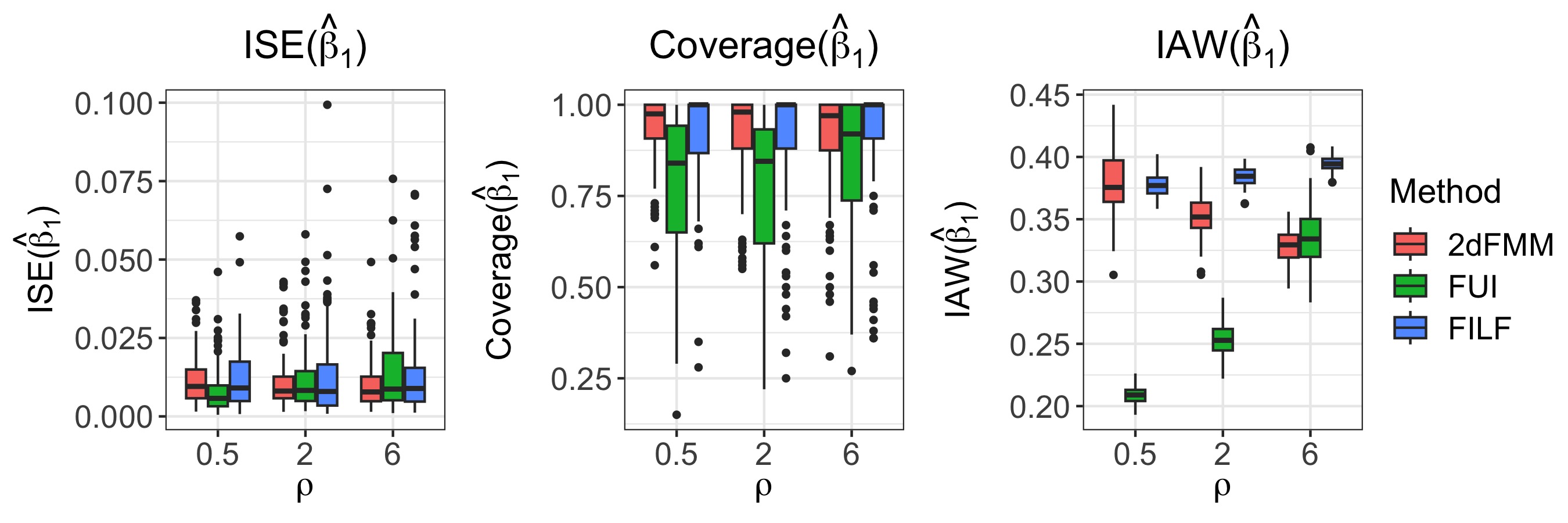}
   \caption{Case (i): the bivariate functional slope $\beta_{1}(s,t)$ under the second scenario (S2) is retained in the true model}
\end{subfigure}
\begin{subfigure}[b]{1\textwidth}
    \includegraphics[width=0.96\textwidth, height=0.32\textwidth]{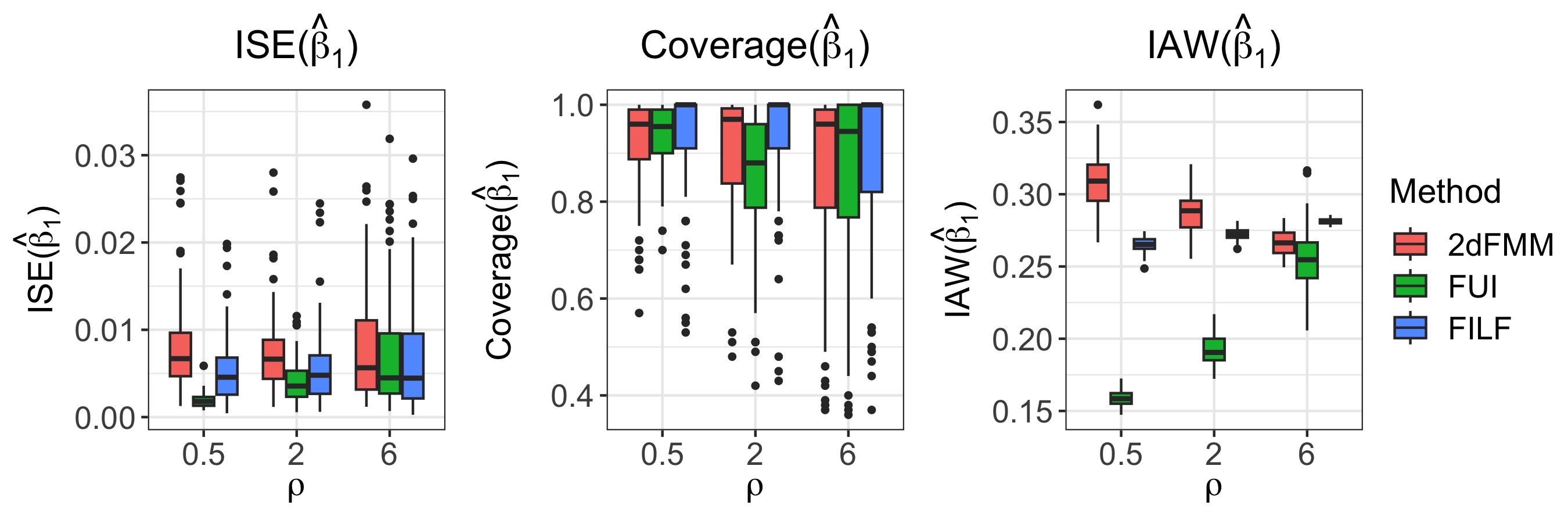}
   \caption{Case (ii): the bivariate functional slope $\beta_{1}(s,t)$ is shrunk to univariate $\beta_{1}(t)$ under the second scenario (S2) in the true model.}
\end{subfigure}
\caption{The comparisons of ISE, coverage probability of 95\% PCB  (Coverage), and IAW for FUI, FILF, and our method with reducing to functional direction with the setting $N = 50$, $R = 20$, and $L=100$. }
\label{fig:simu2_noise_S2S20}
\end{figure}

\begin{figure}[!thb]
\centering
\begin{subfigure}[b]{1\textwidth}
   \includegraphics[width=0.96\textwidth, height=0.32\textwidth]{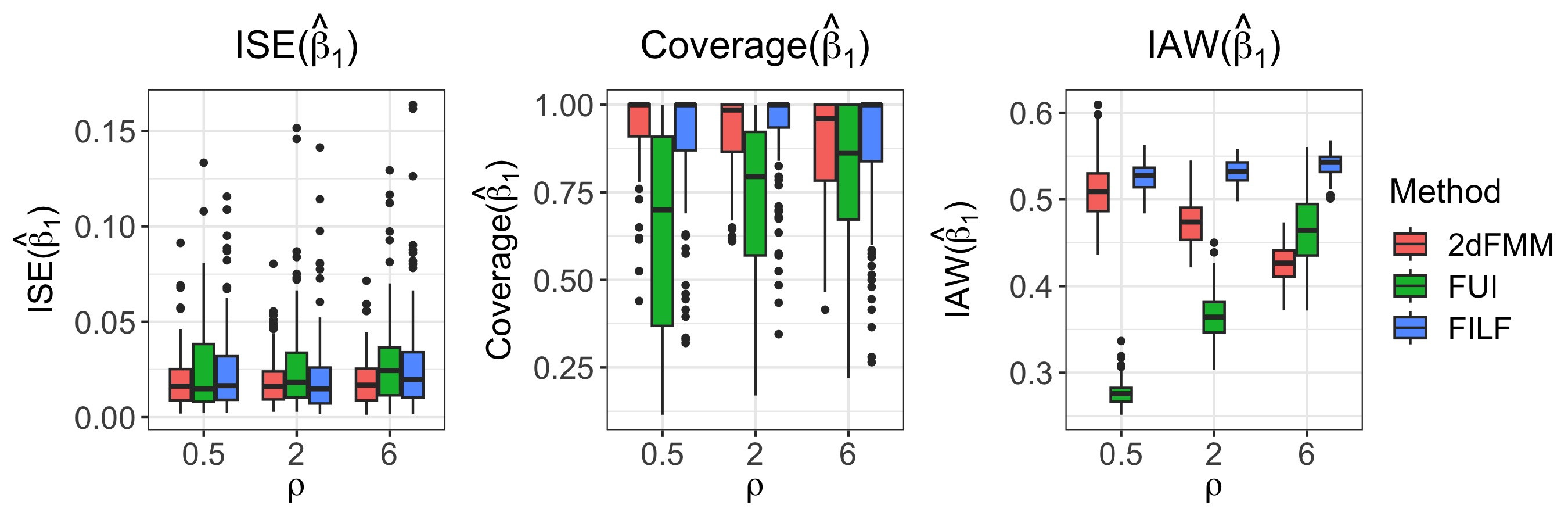}
   \caption{Case (i): the bivariate functional slope $\beta_{1}(s,t)$ under the second scenario (S2) is retained in the true model}
\end{subfigure}
\begin{subfigure}[b]{1\textwidth}
    \includegraphics[width=0.96\textwidth, height=0.32\textwidth]{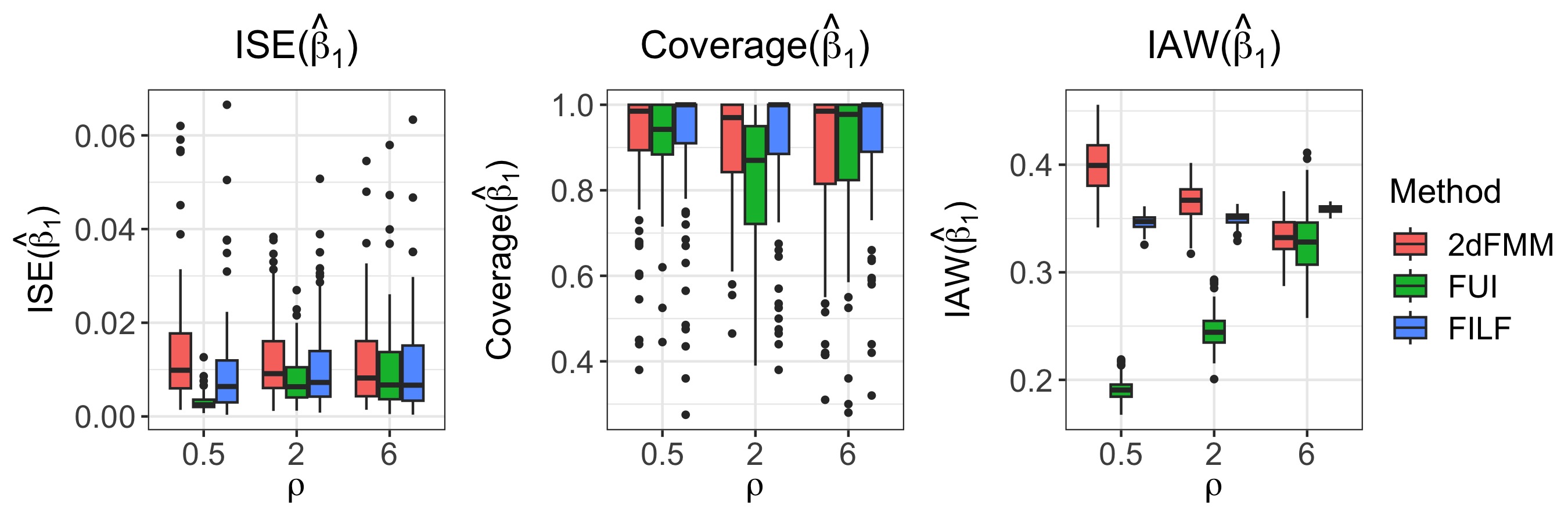}
   \caption{Case (ii): the bivariate functional slope $\beta_{1}(s,t)$ is shrunk to univariate $\beta_{1}(t)$ under the second scenario (S2) in the true model.}
\end{subfigure}
\caption{The comparisons of ISE, coverage probability of 95\% PCB  (Coverage), and IAW for FUI, FILF, and our method with reducing to functional direction with the setting $N = 50$, $R = 10$, and $L=200$. }
\label{fig:simu2_noise_S2L200}
\end{figure}

Table \ref{table:simu_psb&scb} displays the empirical coverage probability of 95\% pointwise confidence bands (PCB) and simultaneous confidence bands (SCB) in S1 (sparse) and S2 (smooth) scenarios, respectively. Figure \ref{fig:simu2_noise_S2N100}-\ref{fig:simu2_noise_S2L200} shows univariate comparative analysis using larger sample generating parameters. 

\begin{table}[!htb]
   \centering
   \scalebox{1}{
\begin{tabular}{c c c c c}
\hline
& & \multicolumn{3}{c}{Sample size ($N$)} \\[0.5ex]
\cline{3-5}
Type & 95\%CB & 50 & 75 & 100 \\[0.5ex]
\hline
\multirow{2}{*}{S1} & PCB & 0.93 (1.45) & 0.93 (1.19) & 0.93 (1.03)\\[0.5ex]
& SCB & 0.95 (1.62) & 0.95 (1.33) & 0.95 (1.14)  \\[1ex]
\multirow{2}{*}{S2} & PCB & 0.97 (1.36) & 0.97 (1.11) & 0.96 (0.96) \\[0.5ex]
& SCB & 0.97 (1.42) & 0.97 (1.15) & 0.97 (0.99) \\[0.5ex]
\hline
& & & Number of grids ($R$) \\[0.5ex]
\cline{3-5}
& & 10 & 15 & 20 \\[0.5ex]
\hline
\multirow{2}{*}{S1} & PCB & 0.93 (1.45) & 0.93 (1.38) & 0.94 (1.39)\\[0.5ex]
& SCB & 0.95 (1.62) & 0.96 (1.58) & 0.97 (1.59) \\[1ex]
\multirow{2}{*}{S2} & PCB & 0.97 (1.36) & 0.97 (1.12) & 0.97 (1.05) \\[0.5ex]
& SCB & 0.97 (1.42) & 0.97 (1.18) & 0.97 (1.09) \\[0.5ex]
\hline
& & & Number of grids ($L$) \\[0.5ex]
\cline{3-5}
& & 100 & 150 & 200 \\[0.5ex]
\hline
\multirow{2}{*}{S1} & PCB & 0.93 (1.45) & 0.93 (1.39) & 0.93 (1.36) \\[0.5ex]
& SCB & 0.95 (1.62) & 0.96 (1.58) & 0.95 (1.53) \\[1ex]
\multirow{2}{*}{S2} & PCB & 0.97 (1.36) & 0.97 (1.36) & 0.96 (1.35) \\[0.5ex]
& SCB &0.97 (1.42) & 0.98 (1.44) & 0.97 (1.42) \\[0.5ex]
\hline
       \end{tabular}}
       \caption{The average empirical coverage of 95\% PCB and SCB under S1 and S2 with varying sample parameters among 100 simulation replicates. The value in the parenthesis corresponds to IAW. The baseline setting is $N = 50$, $R = 10$, and $L=100$. When one parameter is changed, all other sample generating parameters are fixed at their baseline values. } \label{table:simu_psb&scb}
\end{table} 

\clearpage
\newpage

\section{Additional Information for Application}
\label{sec:ai_app}

\begin{figure}[!h]
\centering
\begin{subfigure}[b]{1\textwidth}
    \includegraphics[width=1\textwidth, height=0.555\textwidth]{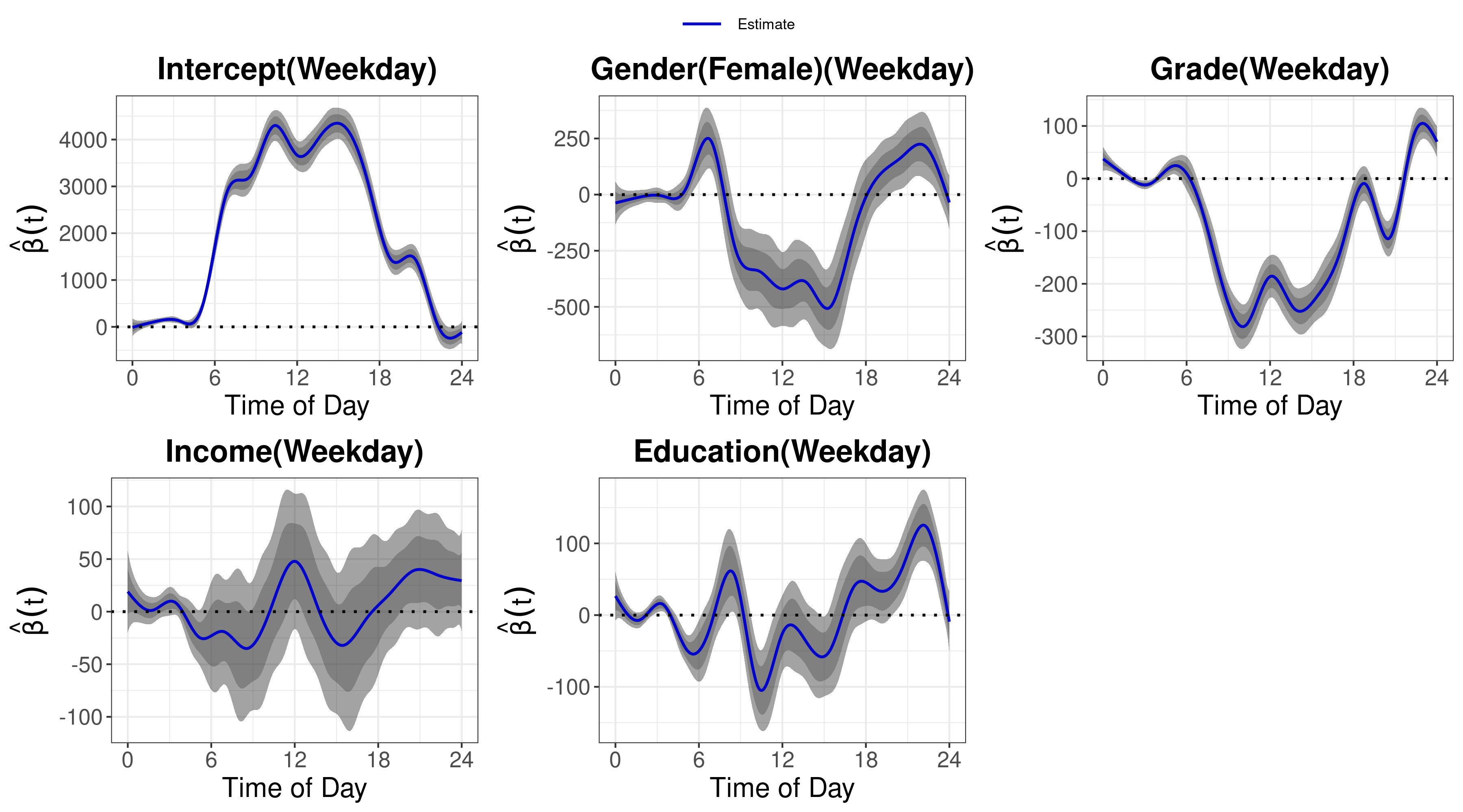}
   \caption{Univariate effect of time of day over weekday.}
\end{subfigure}
\begin{subfigure}[b]{1\textwidth}
    \includegraphics[width=1\textwidth, height=0.555\textwidth]{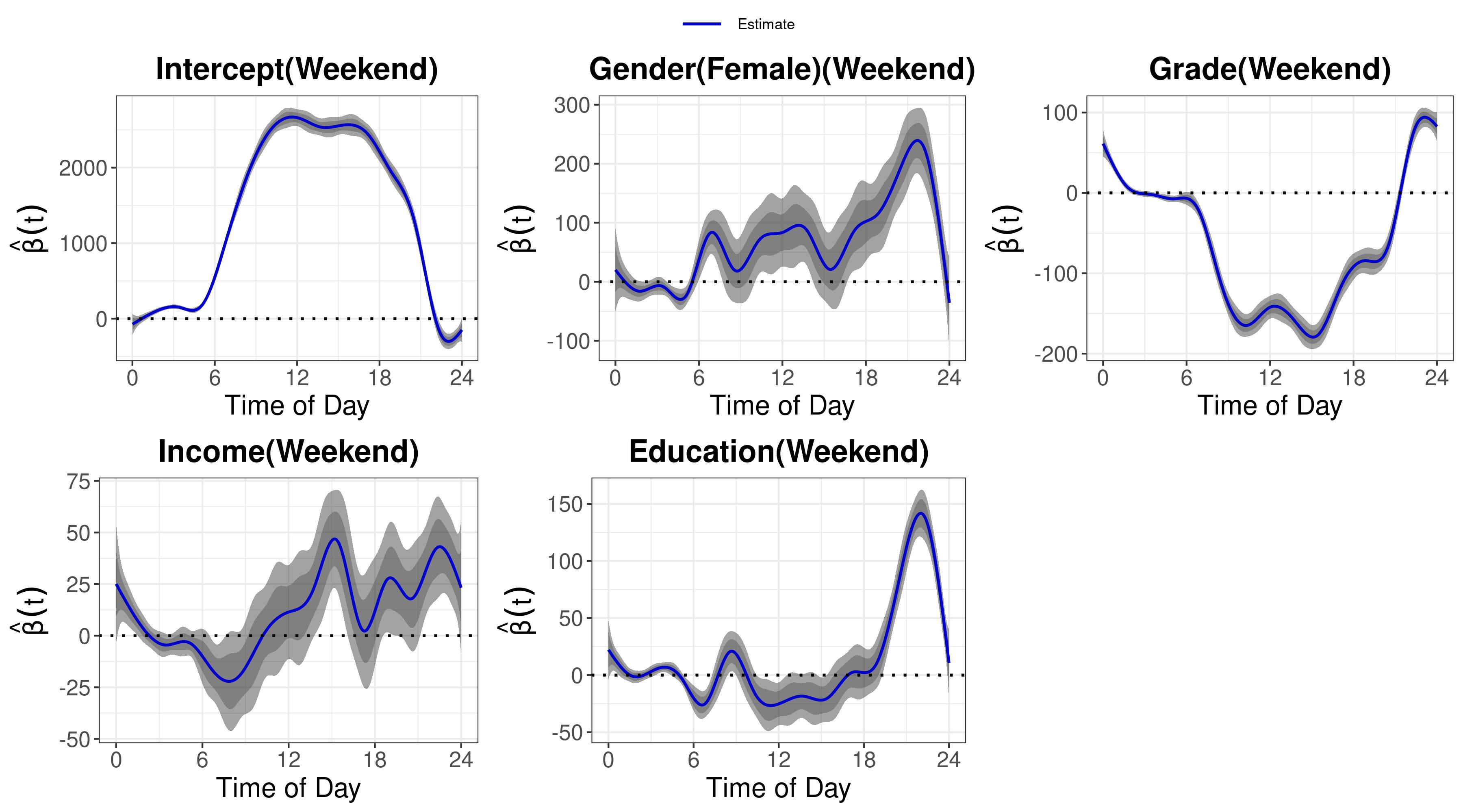}
   \caption{Univariate effect of time of day over weekend.}
\end{subfigure}
\caption{Fixed-effects estimates (dashed blue line), 95\% PCB (dark gray shaded area), and 95\% SCB (light gray shaded area) of physical and intercept, demographic and socioeconomic covariates over the weekday or weekend in the baseline model.}
    \label{fig:real_baseline_weekday_weekend}
\end{figure}

\begin{figure}[!h]
\centering
\begin{subfigure}[b]{1\textwidth}
    \includegraphics[width=1\textwidth, height=0.555\textwidth]{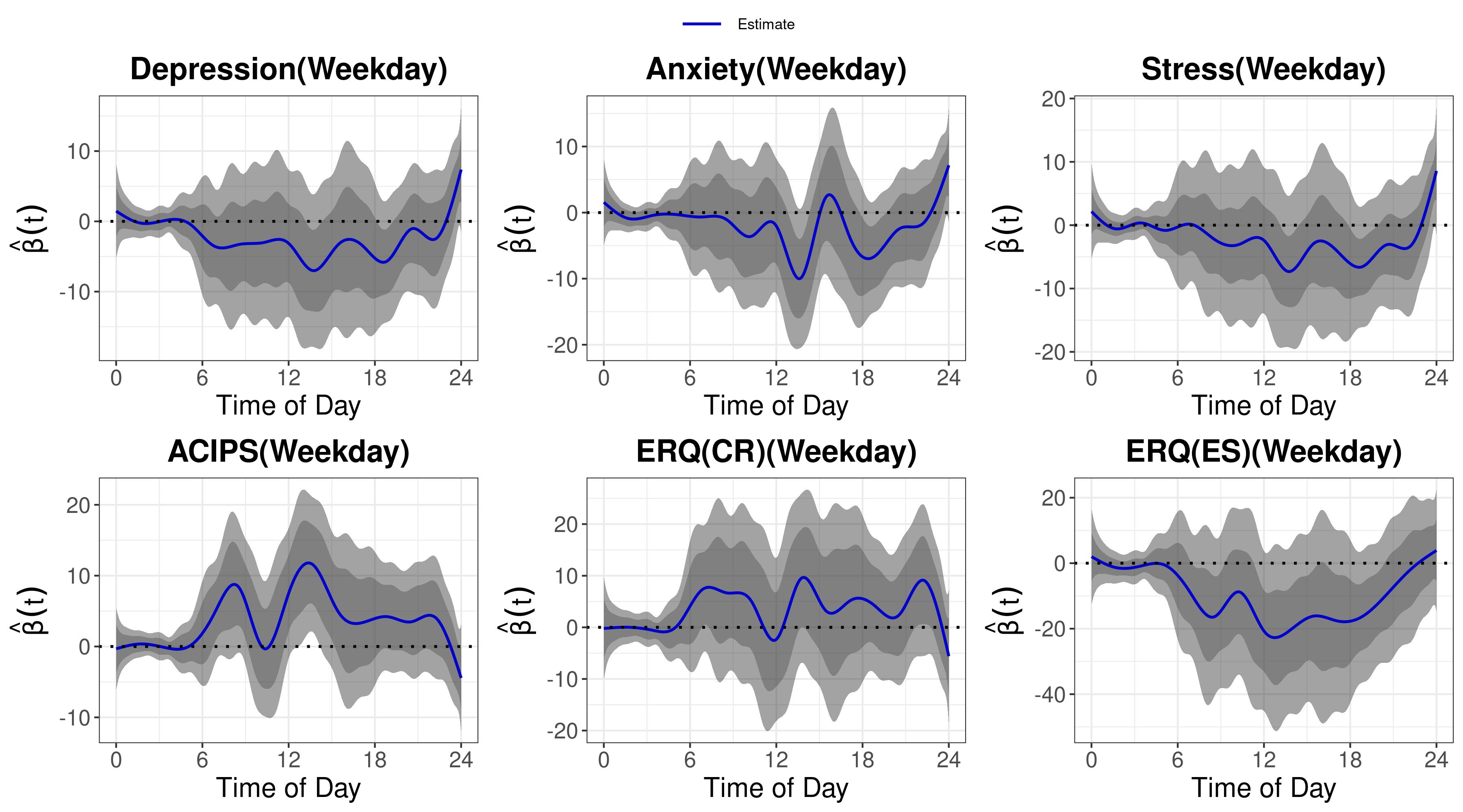}
   \caption{Univariate effect of time of day over weekday.}
\end{subfigure}
\begin{subfigure}[b]{1\textwidth}
    \includegraphics[width=1\textwidth, height=0.555\textwidth]{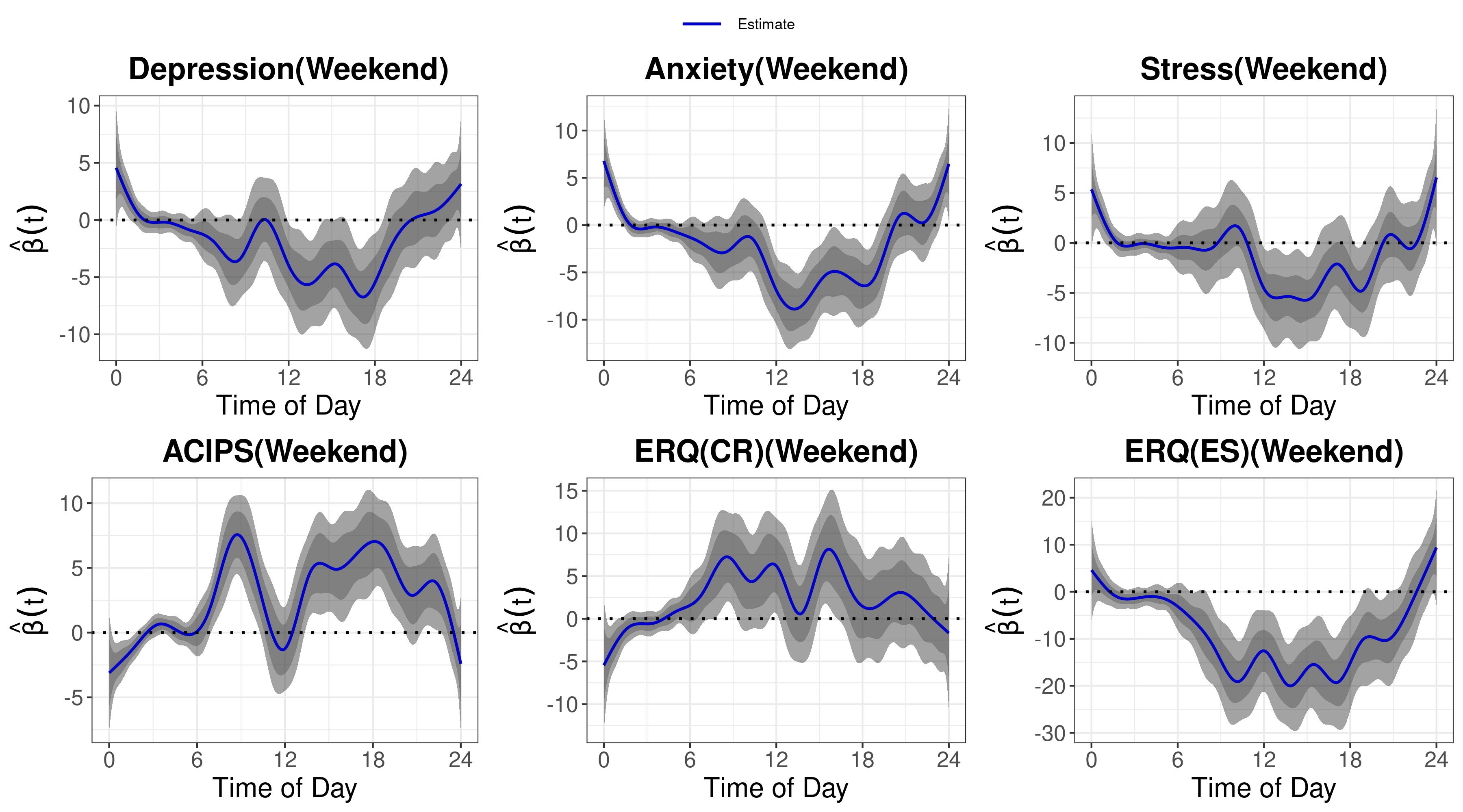}
   \caption{Univariate effect of time of day over weekend.}
\end{subfigure}
\caption{Fixed-effects estimates (dashed blue line), 95\% PCB (dark gray shaded area), and 95\% SCB (light gray shaded area) of mental health outcomes separately added in the baseline model.}
    \label{fig:real_mental_weekday_weekend}
\end{figure}

As in longitudinal functional data analysis, we also assess the fixed effects only over the functional domain. The univariate coefficient function and confidence bands are estimated using the formula of \textit{Remark 3} in Section 2.1. Figure \ref{fig:real_baseline_weekday_weekend} demonstrates the effects of intercept, demographic, and socioeconomic covariates along the time of day for weekdays and weekends, while Figure \ref{fig:real_mental_weekday_weekend} demonstrates the effects of physical and mental health outcomes. 

\clearpage

\bibliographystyle{agsm}
\bibliography{supp}